\documentclass[12pt]{article}

\usepackage{amsmath}
\usepackage{amssymb}
\usepackage{physics}
\usepackage{graphicx}
\usepackage{hyperref}
\usepackage{times}
\usepackage{xcolor}
\usepackage{multirow}
\usepackage{tabularx}
\usepackage{xcolor}
\usepackage{cancel}
\usepackage{geometry}
\usepackage{nicematrix}
\usepackage{bm}
\usepackage{bbm}
\usepackage{comment}
\usepackage{siunitx}
\usepackage{scicite}

\newcommand{\subref}[2]{\hyperref[#1]{\ref*{#1}#2}}
\definecolor{refblue}{HTML}{
2E2E91}
\hypersetup{colorlinks=true, citecolor=refblue, urlcolor=refblue, linkcolor=refblue}

\newenvironment{sciabstract}{%
\begin{quote} \bf}
{\end{quote}}

\title{Coherent evolution of superexchange interaction in seconds long optical clock spectroscopy}

\author
{William R. Milner$^{1\ast}$, Stefan Lannig$^{1}$, Mikhail Mamaev$^{1, 2}$, \\ Lingfeng Yan$^{1}$, Anjun Chu$^{1,2 }$, Ben Lewis$^{1}$, Max N. Frankel$^{1}$, Ross B. Hutson$^{1}$, \\ Ana Maria Rey$^{1, 2}$, and Jun Ye$^{1\ast}$\\
\normalsize{$^{1}$JILA, National Institute of Standards and Technology }\\ 
\normalsize{and University of Colorado, Boulder, CO 80309}\\
\normalsize{$^{2}$Center for Theory of Quantum Matter, University of Colorado, Boulder, CO 80309}\\
\normalsize{$^\ast$Corresponding authors; e-mail: william.milner@colorado.edu, ye@jila.colorado.edu}
}

\date{}
\begin{document}
\maketitle

\begin{sciabstract}
Measurement science now connects strongly with engineering of quantum coherence, many-body states, and entanglement. To scale up the performance of an atomic clock using a degenerate Fermi gas loaded in a three-dimensional optical lattice, we must understand complex many-body Hamiltonians to ensure meaningful gains for metrological applications. In this work, we use a near unity filled Sr 3D lattice to study the effect of a tunable Fermi-Hubbard Hamiltonian. The clock laser introduces a spin-orbit coupling spiral phase and breaks the isotropy of superexchange interactions, changing the Heisenberg spin model into one exhibiting XXZ-type spin anisotropy. By tuning the lattice confinement and applying imaging spectroscopy we map out favorable atomic coherence regimes. With weak transverse confinement, both s- and p-wave interactions contribute to decoherence and atom loss, and their contributions can be balanced. At deep transverse confinement, we directly observe coherent superexchange interactions, tunable via on-site interaction and site-to-site energy shift, on the clock Ramsey fringe contrast over timescales of multiple seconds. This study provides a groundwork for using a 3D optical lattice clock to probe quantum magnetism and spin entanglement.

\end{sciabstract}



Optical lattice clocks are advancing studies of fundamental physics, metrology, and quantum simulation \cite{bothwell2022resolving, kolkowitz2016gravitational, hutson2023observation, boulder2021frequency, nemitz2016frequency, mcgrew2018atomic}. By controlling all external perturbations to the ground and metastable "clock" state, each one of the confined atoms becomes a pristine, two-level system. With clock precision limited fundamentally by quantum projection noise~\cite{itano1993quantum}, a natural approach for improving clock performance is to probe the largest possible number of atoms combined with the longest possible coherence time. However, given a densely packed sample of atoms, we must address outstanding challenges including maintaining a maximum coherence time for clock precision and evaluating systematic effects for clock accuracy. Often it is desirable to minimize atomic interactions to enhance single-particle coherence and control systematic effects. At the same time, as the level of understanding of these interactions becomes more mature and sophisticated, we can engineer a large, coherent spin ensemble with interaction precisely controlled to introduce and optimize quantum coherence, correlation, and entanglement to advance the frontier of quantum metrology~\cite{pedrozo2020entanglement, Robinson2023entanglement, eckner2023realizing,  franke2023quantum}.   



With the ease of geometry tunability, optical lattices provide a versatile platform to confine large numbers of atoms and control their interactions and motion. Over the past two decades, progress in clock precision~\cite{oelker2019demonstration, ludlow2015optical} has been largely advanced by the study and control of interactions in one-dimensional (1D) optical lattice clocks. The corresponding interaction dynamics are well described by a collective spin model ~\cite{martin2013science, zhang2014spectroscopic} that includes both on-site $p$-wave interactions between atoms in different radial modes and off-site $s$-wave interactions. The latter are induced by the spin-orbit coupling (SOC), arising from a difference in wavelength between the clock probe and the lattice laser~\cite{kolkowitz2017spin,Bromley2018}, which lifts the indistinguishability between spin-polarized fermions on neighboring lattice layers along the clock $k$-vector. Systematic exploration of this 1D spin model identified a confinement depth at which the combination of $s$ and $p$-wave interactions suppressed detrimental mean-field density shifts~\cite{bothwell2022resolving,aeppli2022hamiltonian}. These advances based on precise experimental control motivates quantum simulation investigations of the comparatively less-studied three-dimensional (3D) lattice spin model~\cite{campbell2017fermi}. 

In a 3D lattice filled with a degenerate Fermi gas of spin-polarized $^{87}$Sr atoms in the motional ground state~\cite{sonderhouse2020thermodynamics}, the system can be modelled with the Fermi-Hubbard Hamitonian where ground and excited state atoms on the same lattice site interact via the Hubbard interaction parameter $U$, and motion is captured by a tunneling parameter $t$. In the unity filled limit a Mott-insulating regime emerges  at $U \gg t$,  atomic motion is restricted, and atoms interact only via virtual second order tunneling processes that induce spin-exchange couplings between nearest neighbour atomic spins known as superexchange \cite{duan2003controlling, lewenstein2007ultracold, trotzky2008time}. The physics of superexchange is central in describing magnetic phenomena such as antiferromagnetism ~\cite{manousakis1991spin, auerbach2012interacting} and is believed to play a role in superconductivity~\cite{lee2006doping}. Several ultracold atom experiments have employed optical lattices to explore low-temperature bosonic ferromagnetic and fermionic antiferromagnetic correlations induced by superexchange~\cite{greif2013short, hart2015observation, boll2016spin, cheuk2016observation, mazurenko2017cold, takahashi2005thermodynamics, taie2022observation, gall2021competing}, as well as some non-equilibrium superexchange-driven quantum dynamics in local density probes~\cite{brown2015two, jepsen2020spin, sun2021realization}. With the goal of  achieving optimal and scalable clock performance at a unity filled 3D lattice, understanding and controlling the effects of superexchange on collective spin dynamics becomes necessary~\cite{campbell2017fermi, goban2018emergence}. 
The current work employing seconds long Ramsey spectroscopy on tens of thousands of atoms directly probes the coherent nature of superexchange interaction, thus strengthening our understanding of interaction regimes that are favorable for robust quantum coherence and entanglement. 

In the current experiment we independently vary the lattice confinement to explore the 1D and 3D lattice spin models, including the crossover between the two regimes for the first time.  To do so we load a degenerate Fermi gas of $^{87}$Sr atoms into a 3D lattice with tunable confinement, allowing us to vary the interaction strength and tunneling rates. The interaction effects on spin coherence between the ground and metastable clock state are directly recorded on Ramsey fringes. In a vertical 1D lattice, we achieve coherence times of $\sim$20 s when minimizing the contribution of $s$ and $p$-wave interactions. As a weak transverse confinement is turned on, $s$-wave interactions are increased by orders of magnitude and very fast dephasing is observed.  At deep transverse confinement, favorable coherence times are partially recovered, and coherent superexchange interactions are manifested directly in oscillations of the Ramsey fringe contrast persisting over a timescale of multiple seconds. These experimental observations are well captured by an anisotropic lattice spin model (XXZ plus antisymmetric exchange terms), which breaks the Heisenberg SU(2) symmetry of the Fermi-Hubbard physics due to the spin-orbital coupling phase~\cite{kolkowitz2017spin,Bromley2018,mamaev2021tunable, jepsen2022long}. Realization of anisotropic spin interactions in controlled cold atomic systems has only recently seen exploration, and is highly relevant to studies of spin magnetism~\cite{takahashi2005thermodynamics} and transport~\cite{bertini2021finite}. In clocks, such interactions can also be directly employed for the generation of large scale quantum entanglement over the entire 3D lattice system~\cite{he2019engineering, yanes2022one, mamaev2023spin}.

\begin{figure*}[hbtp]
    \centering
    \includegraphics[width=5.5in]{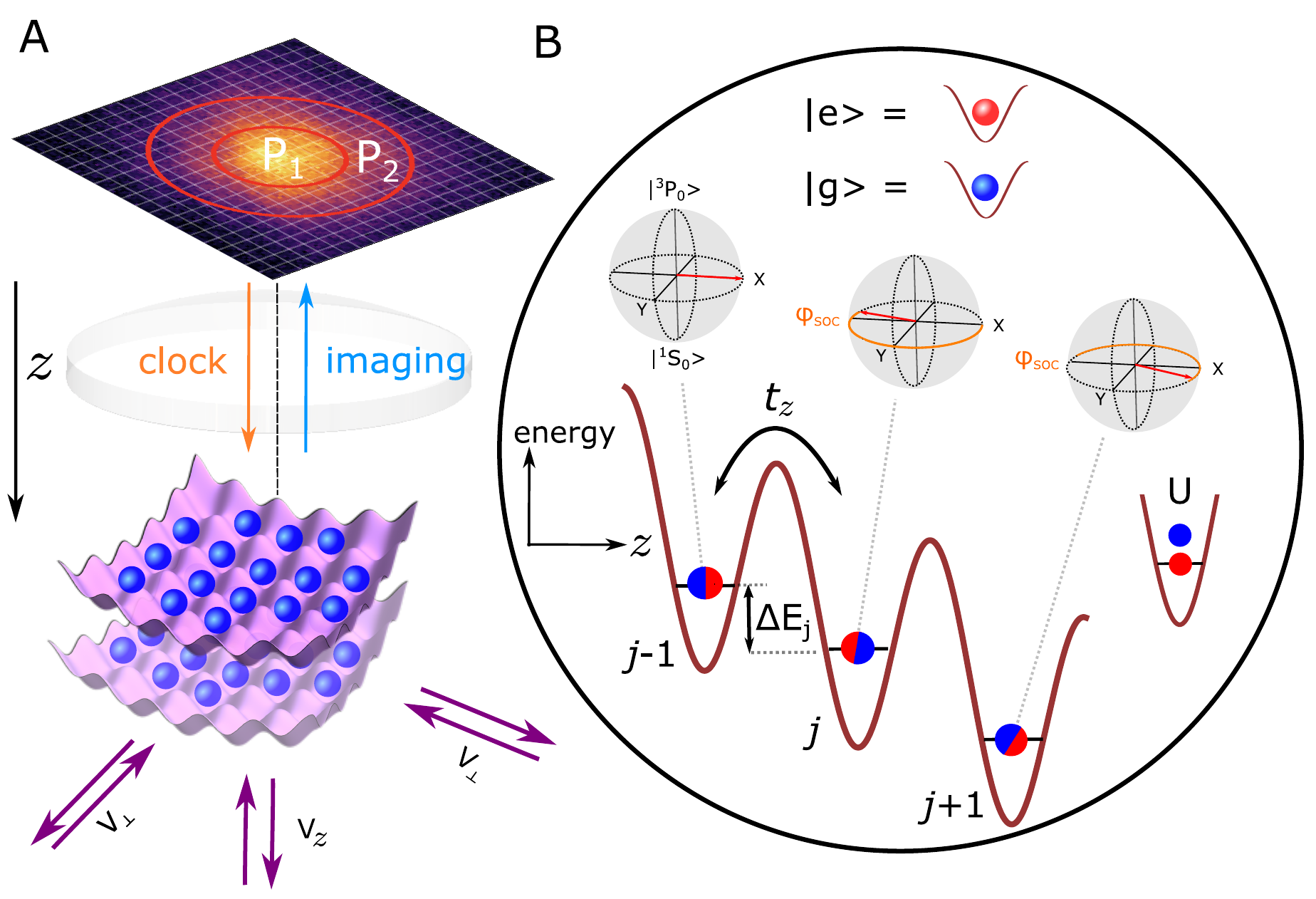}
    \caption{
	    \textbf{Experimental setup and interaction model}.  \textbf{A,} Ultracold fermions are confined in the ground band of a three-dimensional optical lattice with tunable confinement. Lattice depths can be independently varied by changing the optical power of retro-reflected beams in the transverse $V_{\perp}$ or vertical direction $V_{z}$. \textit{In situ} imaging allows to spatially resolve interactions and dephasing via imaging spectroscopy \cite{marti2018imaging}. \textbf{B,}  Dynamics are described via the Fermi-Hubbard model with tunneling $t_{z}$, interaction energy $U$, and a site-to-site energy shift $\Delta E_{j}$ from the lattice Gaussian confinement. Atoms along the $z$ axis on sites indexed $j -1, j$  are initialized in a superposition state of the ground state $\ket{g = \, ^{1}S_{0}}$ and the metastable electronic state ("clock" state) $\ket{e = \, ^{3}P_{0}}$, where the clock laser imprints local phase shift $\varphi$ due to spin-orbit coupling. Dephasing of the coherence is proportion to an effective superexchange rate: $4t_{z}^2U/(U^2 - \Delta E_{j}^2)$}
    \label{fig:fig1}
\end{figure*}

The experimental schematic is depicted in Fig.~\subref{fig:fig1}{A}. After evaporation, we confine the atoms in a retroreflected, cubic lattice operating at the magic wavelength of $\lambda_{magic}$ = $813$ nm with lattice constant $a \approx 407$ nm~\cite{campbell2017fermi}.
Beginning with a nuclear-spin polarized Fermi gas with a temperature $T/T_{F} \approx 0.2$, the atoms are adiabatically loaded into the ground band of the 3D lattice \cite{sonderhouse2020thermodynamics, campbell2017fermi}.  In the deep lattice, the initial state is nearly a band insulator with a peak filling of one atom per lattice site \cite{milner2023high, hutson2023observation}. 
The lattice depth $\bigl(V_{\perp}\bigr)$ of the transverse (horizontal with respect to gravity) confinement is tuned independently from the depth of the vertical confinement $\bigl(V_{z}\bigr)$ by adjusting the optical power in the corresponding lattice beams.
Our two-level spin system is established between the ground $^{1}S_{0}$ $\bigl(\ket{g}\bigr)$ and metastable electronic "clock" state $^{3}P_{0}$ $\bigl(\ket{e}\bigr)$. We coherently drive the clock transition $\ket{g, m_{F} = -9/2 } \leftrightarrow \ket{e, m_{F} = -9/2 }$ at $\lambda_{clk} \approx 698$ nm with a vertical laser beam using an optical local oscillator locked to an ultrastable silicon cavity~\cite{matei20171}.

After loading the lattice, we put the atoms into a superposition of $\ket{g}$ and $\ket{e}$ and perform Ramsey spectroscopy. For detection, \textit{in situ} absorption imaging along the vertical direction is employed and approximately 100 photons per atom are scattered over a $1$ $\mu$s pulse duration with minimal blurring compared to the diffraction-limited point-spread function of $1.3$ $\mu$m \cite{milner2023high, hutson2023observation}. Two images of the ground and clock state atoms, their numbers denoted $N_{g}$ and $N_{e}$, are taken to determine the excitation fraction $p_{e} = N_{e}/(N_{e} + N_{g})$. For a chosen region-of-interest $P_{A}$ of our imaged density distribution, we record the local excitation fraction $p^{A}_{e} = N^{A}_{e}/(N^{A}_{e} + N^{A}_{g})$. This is shown in Fig.~\subref{fig:fig1}{A}, where the excitation fractions are evaluated in spatially separate regions $P_{1}$ and $P_{2}$ to determine both the Ramsey fringe contrast and relative atomic coherence using imaging spectroscopy \cite{marti2018imaging}.

During the Ramsey interrogation time the atoms interact via the Fermi-Hubbard model presented in Fig.~\subref{fig:fig1}{B} \cite{murmann2015two}. The on-site interaction $U = \frac{4 \pi \hbar^{2}}{m} a_{eg-} 
 \int |W(\bold{r})|^{4} d^{3}\bold{r}$ is determined by the anti-symmetric scattering length $a_{eg-} = 69.1(0.9) a_{B}$ \cite{goban2018emergence, zhang2014spectroscopic}  and the 3D, single-particle Wannier function $W(\bold{r})$ is determined by the lattice confinement. Along the vertical direction, $\textit{z}$, the atoms on neighboring sites are coupled with the tunneling rate $t_z$, and also experience both the linear gravitational  potential  and the confinement from the Gaussian transverse lattice beams, leading to an energy offset $\Delta E_{j}$ between adjacent vertical lattice planes indexed by $j$ \cite{lemonde2005optical}.   The clock laser is also  launched along the vertical direction  imprinting a spin-orbit-coupling (SOC) phase $\varphi = 2 \pi a / \lambda_{clk} \approx 7 \pi /6$ between neighboring  vertical lattice planes \cite{kolkowitz2017spin, hutson2019engineering} as depicted in Fig.~\subref{fig:fig1}{B}. 
  


The superexchange oscillations, observed in  the deep 3D confinement regime of our experiment, can be understood from a simple double-well model describing two atoms (spin $s = \frac{1}{2}$) on two adjacent lattice sites $j=0,1$ along the $z$ lattice direction. The Ramsey spectroscopy protocol initializes the atoms in a superposition state $\ket{\psi_{\textrm{init}}}$ = $(\ket{g}_{0} +\ket{e}_{0}) / \sqrt{2}$ $\otimes$ $(e^{-i\varphi/2}\ket{g}_{1}+e^{+i\varphi/2}\ket{e}_{1}) / \sqrt{2}$. Crucially, due to the site-dependent, spin-orbit coupling phase $\varphi$, aside from a global phase this initial state is an admixture of the spin triplet and singlet states, with $\ket{\psi_{\textrm{init}}} \sim e^{-i \varphi/2}\ket{g,g} + e^{i \varphi/2}\ket{e,e}+\cos(\varphi/2)(\ket{g,e}+\ket{e,g}) + i \sin(\varphi/2) (\ket{g,e}-\ket{e,g})$. At half-filling and in the strongly interacting limit, $U\gg t_z$, superexchange interactions arising between neighboring spins,  $J_{\mathrm{SE}} \hat{\tilde{\bold{s}}}_{0} \cdot \hat{\tilde{\bold{s}}}_{1}$, introduce an energy shift for the singlet state, which translates to a phase difference $J_{\mathrm{SE}} T$ compared to the triplet states during the coherent evolution time $T$. Here $\hat{\tilde{s}}_{j}^{\alpha}$ for $\alpha \in \{ X, Y, Z \}$ refers to spin-1/2 matrices describing atoms on sites $j$ in the lab frame.


More formally,  we rotate into a ``spiral'' frame where the initial state is uniform (all atoms in the same superposition state) and the site-dependent laser phase $\varphi$ is absorbed into the spin operators across the lattice, $\hat{s}^{\pm}_j=\hat{\tilde{s}}^{\pm}_je^{\pm ij\varphi}$, $\hat{s}^{Z}_j=\hat{\tilde{s}}^{Z}_j$.  Thus, we obtain a superexchange spin Hamiltonian in the spiral frame 
\begin{equation}
\hat{H}_{\mathrm{SE}} =  \sum \limits_{j} J_{\mathrm{SE}}(j) \Big[ \frac{1}{2}\left(e^{i \varphi}\hat{s}^{+}_{j}\hat{s}^{-}_{j + 1} + H.c.\right)  + \hat{s}^{Z}_{j}\hat{s}^{Z}_{j + 1}\Big].
\end{equation}
The superexchange interaction strength is $J_{\mathrm{SE}}(j) = 4t_{z}^{2}U/(U^{2} - \Delta E_{j}^{2})$, which is inhomogeneous due to the local potential difference between adjacent sites $\Delta E_{j}$, including gravity and the lattice Gaussian confinement. Furthermore, the spiral phase makes this spin Hamiltonian go beyond conventional superexchange interactions in optical lattices, as it exhibits exchange-symmetric XXZ-style anisotropy and an antisymmetric spin exchange term~\cite{supplement}.  Observables such as atomic coherence reveal collective quantum dynamics on timescales of the averaged $\bar{J}_{\mathrm{SE}}$ over the ensemble, which is tuned by controlling the inhomogeneity and the lattice depth.

The above theoretical description is valid in the regime $V_{z}\ll V_{\perp}$, for which the system acts as individual vertical tubes and each site with an atom acts as a spin-$1/2$ particle. Prior work~\cite{aeppli2022hamiltonian} has also shown that in the 1D lattice confinement along $z$ ($V_{\perp}=0$), each lattice site holds many atoms. 
In this 1D limit, on-site interactions favor spin alignment between  atoms, locking them into large collective spins of Wannier-Stark level $n$ along gravity ,  $\hat{S}^\alpha_n$,
whose dynamics is described by the same type of spin Hamiltonian as superexchange but with modified  couplings and an additional onsite term. 
$\hat{H}_{\mathrm{LS}}=\hat{H}_{\mathrm{on-site}}+\hat{H}_{\mathrm{off-site}}$. Here, $\hat{H}_{\mathrm{on-site}}\sim \sum_n \hat{ S}^Z_n\hat{S}^Z_n$ describes the on-site $p$-wave interactions (see Fig.~\subref{fig:fig3}{A}), and  $\hat{H}_{\mathrm{off-site}}$ includes  the off-site $s$-wave interactions  and takes the same form as $\hat{H}_{\mathrm{SE}}$ by replacing the spin-$1/2$ operators with large-spin operators. In this work, we bridge these two regimes by varying the transverse lattice confinement $V_{\perp}$. We extend the theoretical description of Ref.~\cite{aeppli2022hamiltonian} to the regime $V_{z}\gg V_{\perp}$, where in each pancake the weak transverse lattice defines a new set of transverse eigenmodes with renormalized  spin couplings \cite{supplement}. 

 



\begin{figure*}[hbtp]
    \centering
    \includegraphics[width=5.5in]{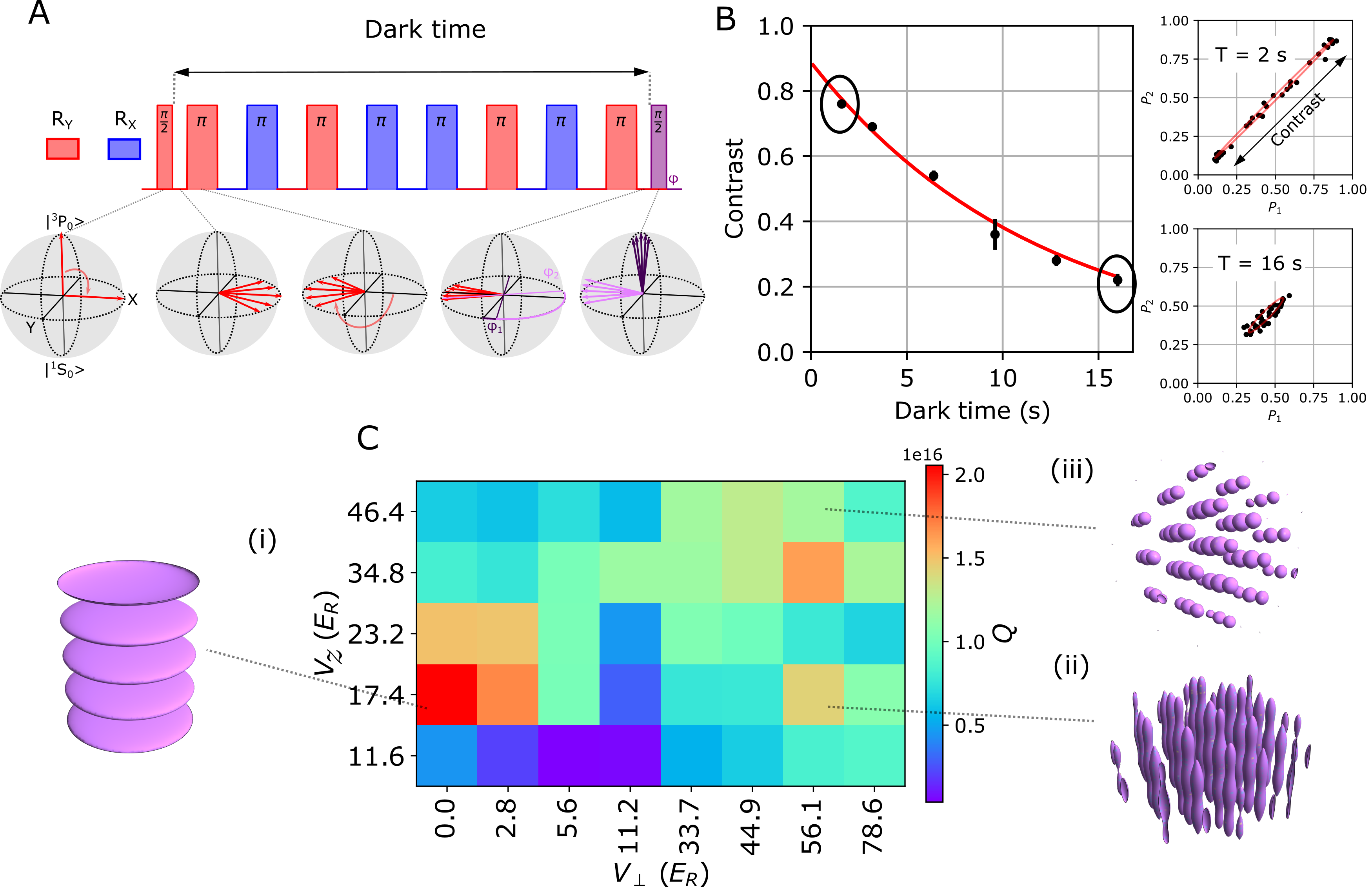}
    \caption{\textbf{Coherence time measurement}.  
	    \textbf{A},  Ramsey spectroscopy is employed to study the coherence time.  An XY8 pulse sequence is used to mitigate single-particle dephasing. The dephasing and rephasing of individual spins is depicted on the Bloch sphere during the echo sequence. For the final $\pi/2$ pulse two choices of the randomized phase $\varphi_{1,2}$ are shown (light and dark purple) to illustrate the spread of resulting excitation fractions in individual realizations.
        \textbf{B}, To determine the coherence time $T_2$, the contrast decay is fit to a stretched exponential $C(T) = C_{0} e^{-(T/T_{2})^{\alpha}}$ as a function of dark time $T$. The contrast is determined via parametric plots of  excitation fractions in regions $P_{1}$ and $P_{2}$ of the ensemble as depicted in Fig. 1. Error bars are $1 \sigma$ (standard deviation)  obtained from jackknifing.
            \textbf{C}, The quality factor $Q = \pi C_{0}T_{2}\nu$ where $\nu \approx 429$ THz is plotted over a wide range of transverse and vertical confinement. Two candidate regimes are identified to investigate further. The weak or zero transverse confinement regime (i), where the longest optical lattice clock $T_{2}$ times have been reported \cite{bothwell2022resolving}. Regime (ii), where fast initial contrast decay is observed due to superexchange interactions. The deep 3D lattice regime (iii) was studied on this platform in \cite{hutson2019engineering} where the coherence time is limited by Raman scattering of lattice photons.  }
    \label{fig:fig2}
\end{figure*}

\begin{figure*}[hbtp]
    \centering
    \includegraphics[width=5.5in]{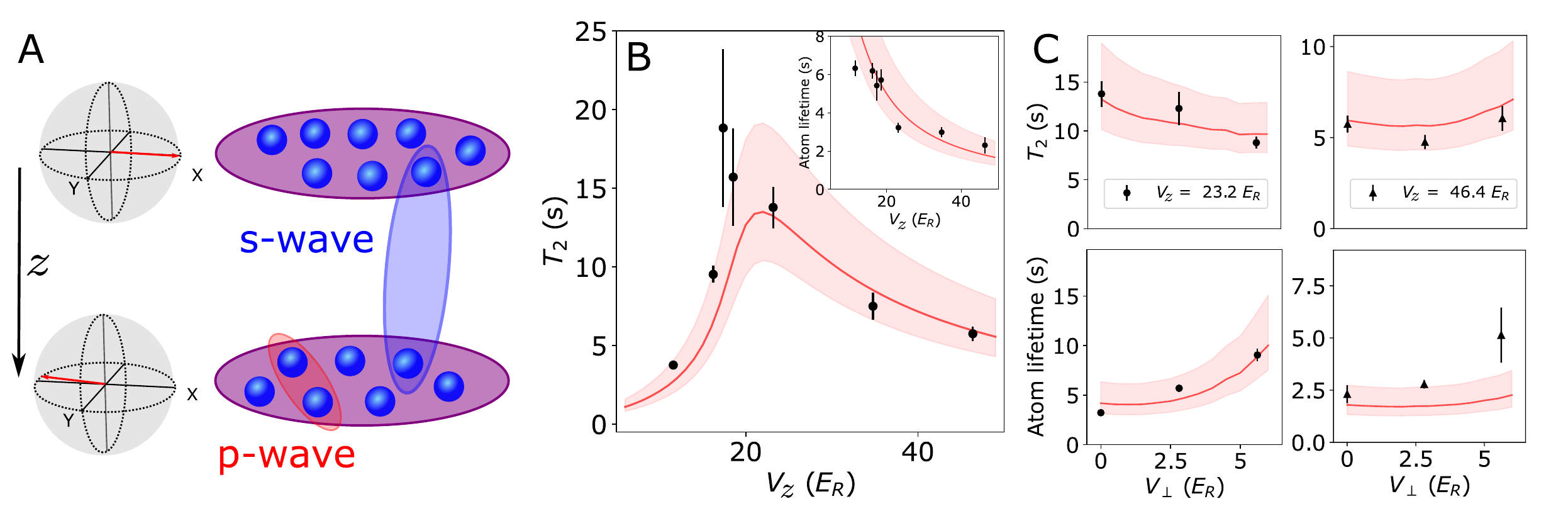}
    \caption{
	    \textbf{Weak transverse confinement regime}. \textbf{A},  In the weak transverse confinement regime, both off-site $s$-wave interactions, induced by the SOC phase between lattice sites, and on-site $p$-wave interactions between atoms contribute to dephasing  \cite{aeppli2022hamiltonian}. Their strength is controlled by the vertical confinement $V_{z}$ and transverse confinement $V_{\perp}$, strongly influencing the observed coherence time $T_{2}$. 
            \textbf{B}, $T_{2}$ is measured without transverse confinement ($V_{\perp}$ = 0).  In the inset the atom lifetime $\tau$, limited by inelastic p-wave loss, is plotted as a function of $V_{z}$~\cite{supplement}. Theory modeling Ramsey contrast decay based on the 1D spin Hamiltonian using experimental measured parameters is overlaid in red. The error bands are based on the uncertanties of the experimental parameters~\cite{supplement}. Error bars are $1 \sigma$ (standard deviation) uncertainty of the fitted $T_{2}$ and $\tau$ values. 
        \textbf{C}, A weak transverse confinement $V_{\perp}$ is applied. This leads to increased $\tau$, as well as a reduction of $T_2$ at intermediate $V_{z} = 23.2 E_{R}$ and a enhancement of $T_2$ at deep $V_{z} = 46.4 E_{R}$.} 
    \label{fig:fig3}
\end{figure*}

To evaluate atomic coherence that is related to clock performance at different lattice confinement, we measure the Ramsey fringe contrast for varying dark time $T$. An XY8 sequence consisting of eight $\pi$ pulses along the two orthogonal rotation axes in the equatorial place of the Bloch sphere is used to remove single particle dephasing as depicted in Fig.~\subref{fig:fig2}{A} \cite{gullion1990new, li2023tunable}. To decouple the atomic coherence measurement from the finite atom-light coherence time ($\sim$3 s) \cite{matei20171}, the phase of the final Ramsey $\pi$/2 pulse is randomized.  Parametric plots of the excitation fractions from concentric regions $P_{1}$ and $P_{2}$ ($P_{1}$  $\textless \;  6 \mu$m and $6  \; \mu$m $\textless$ $P_{2}$ $\textless  \;  12 \mu$m with respect to the trap center) are used to determine the contrast as shown in Fig.~\subref{fig:fig2}{B}. These parametric plots show ellipses, where a maximum likelihood estimator determines the ellipse contrast and jackknifing is used to extract 1$\sigma$ (standard deviation) errorbars for all Ramsey contrast measurements~\cite{marti2018imaging}. The system is sufficiently homogeneous in the spatial regions $P_{1}$ and $P_{2}$ that the contrast $C$ is approximately the same~\cite{supplement}. No statistically significant phase shift between $P_{1}$ and $P_{2}$ is measured, indicating that the XY8 pulse sequence largely removes any spatially varying frequency shift. 

As a function of dark time $T$, a stretched exponential function $C_{0} e^{-(T/T_{2})^{\alpha}}$ is fit to the Ramsey contrast to extract a $T_{2}$ coherence time for $T \; \textgreater \; 1$ s. For $V_{\perp} = 0$, we expect intra-site, all-to-all p-wave interactions to lead to Gaussian decoherence. We extract a single value $\alpha = 1.38$ by minimizing the combined $\chi^{2}$ for all measurements for $V_{\perp} = 0$ . For all other measurements with $V_{\perp} > 0$, we set $\alpha = 1$ when fitting $T_{2}$. The extracted quality factor $Q = \pi C_{0}T_{2}\nu$ is plotted in Fig.~\subref{fig:fig2}{C}, where $\nu$ is the clock transition frequency $\approx 429$ THz.  We identify two interesting regimes to investigate further: (1) In the 1D lattice regime with no transverse confinement the longest coherence times are observed; (2) With deep transverse confinement where the average $\bar{J}_{\mathrm{SE}} / h \gtrsim 1$ Hz, coherent superexchange dynamics are observed on the Ramsey fringe contrast over a timescale of seconds. As previously reported~\cite{hutson2019engineering}, the deep 3D lattice regime (3) where $\bar{J}_{\mathrm{SE}} / h \ll 1$ Hz reveals a limit on the coherence time primarily due to Raman scattering of lattice photons on $\ket{e}$ atoms. The dark times in this study ($ T \textless$ $16$ s) are short compared to both the $^{1}S_{0}$ lattice lifetime and vacuum lifetime~\cite{supplement}.


Intrigued by the results in Fig.~\subref{fig:fig2}{C}, we compare the 1D and 3D confinement regimes (1) and (2). In 1D ($V_{\perp}=0$), both on-site $p$-wave and off-site $s$-wave interactions contribute to the contrast decay (see Fig.~\subref{fig:fig3}{A}). The observed $T_2$ coherence time and atom lifetime are plotted as a function of $V_z$ in Fig.~\subref{fig:fig3}{B}.
Varying $V_z$ provides two distinct regimes to probe the physics of contrast decay. At large $V_{z}$, atoms become localized in Wannier orbitals along the $z$-lattice and interact predominantly via on-site Ising-type $p$-wave interactions that contribute to slow contrast decay with $T_2\sim 1/\sqrt{N_{s}}$ ($N_s$ is the atom number per pancake), as observed in previous studies \cite{martin2013science}. As $V_z$ decreases, the reduced $p$-wave interaction leads to slower decoherence rate.  However, the Wannier-Stark states become increasingly delocalized along $z$ and atoms experience  progressively stronger off-site $s$-wave interactions. The interplay between $s$-wave and $p$-wave interactions leads to spin wave instabilities that contribute to fast contrast decay with $T_2\sim 1/N_s$~\cite{supplement}. With increasing $s$-wave interaction strength, this instability rate increases as $V_z$ decreases. The crossover between these two mechanisms occurs around $V_z=17.4E_{R}$, where $E_{R} = h^{2}/8ma^{2} \approx h \times 3.5 $ kHz is the lattice photon recoil energy, with a correspondingly longest coherence time of $19(5)$ s. 
While experiment and theory largely agree with each other, the discrepancy at long coherence times could arise from unexpected reduction of the $s$-wave interaction strength from re-thermalization processes neglected in the theory. 
The 1D lattice employed in this study operates with a much higher density than previous studies~\cite{bothwell2022resolving, aeppli2022hamiltonian}. Thus, the atom lifetimes (see Fig.~\subref{fig:fig3}{B} inset), limited by inelastic $p$-wave loss are correspondingly much shorter~\cite{bothwell2022resolving}. 

Upon introduction of a weak transverse confinement ($V_{\perp}\ll V_{z}$), the increasing localization of the transverse modes in the $x$-$y$ plane leads to enhancement of $s$-wave interactions. Additionally, due to decreased overlap of transverse modes, $p$-wave interactions are suppressed, which in turn substantially improves atom lifetimes as shown in Fig.~\subref{fig:fig3}{C}. Meanwhile, different trends in the coherence time are observed between the intermediate $V_z=23.2E_R$ and deep $V_z=46.4E_R$ lattices. 
For $V_z=23.2E_R$, the weak transverse confinement increases $s$-wave interactions within pancakes, enhancing  the population of unstable spin wave modes, and a subsequent decrease of $T_2$. For $V_z=46.4E_R$, the system remains in the quasi-stable Ising dominated regime and $T_2$ increases as $p$-wave interactions decrease.



\begin{figure*}[hbtp]
    \centering
    \includegraphics[width=5in]{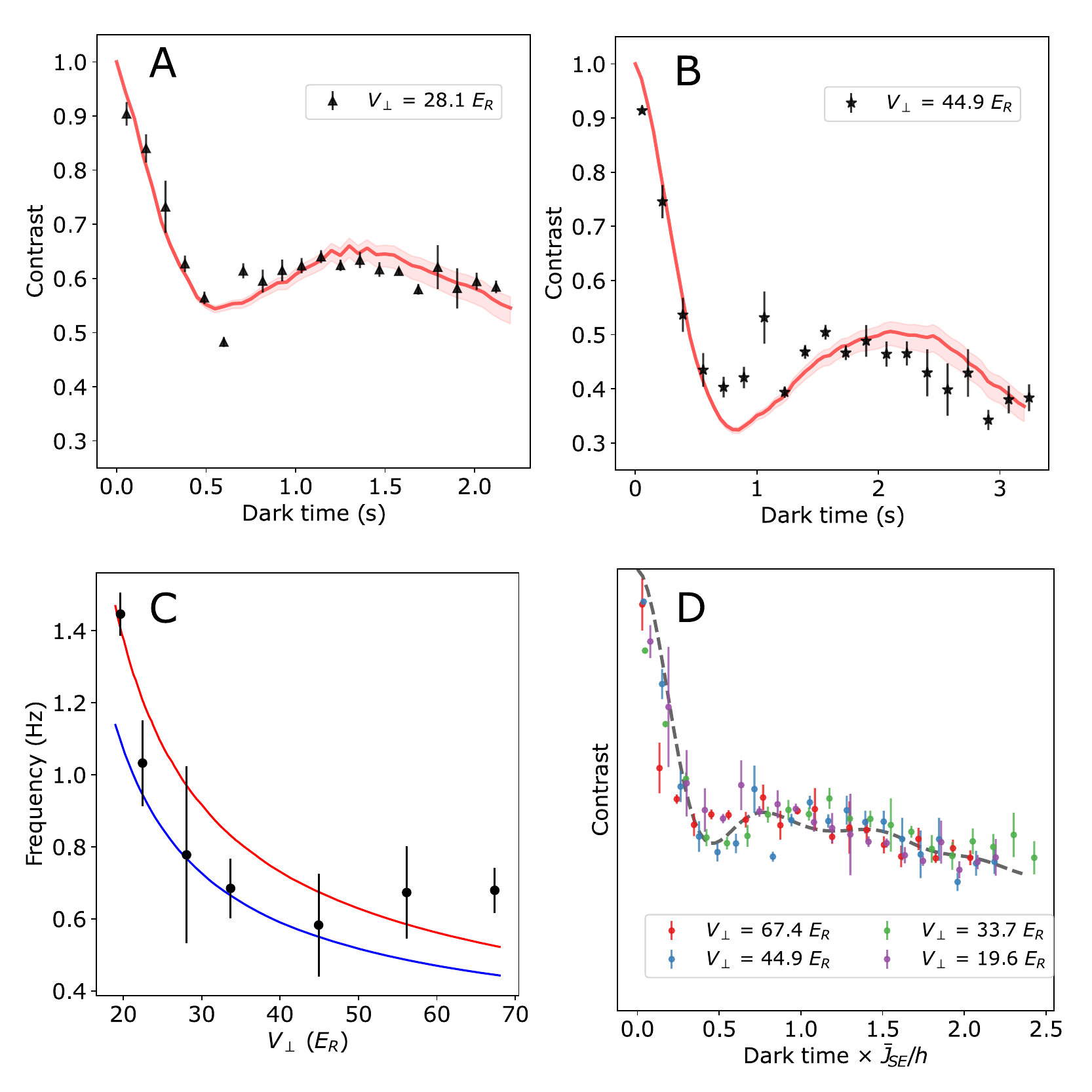}
    \caption{
	    \textbf{Observing superexchange interactions}. 
      Ramsey contrast decay is studied in a 3D lattice at fixed $V_{z}$ = 17.4 $E_{R}$ and thus $t_{z}$, while $V_{\perp}$ is varied between approximately 70 and 20 $E_{R}$ primarily modifying $U$.  Decay curves at $V_{\perp}$ =  28.1 $E_{R}$ \textbf{A}, and  44.9 $E_{R}$ \textbf{B}, are plotted. Error bars are $1 \sigma$ (standard deviation). Red lines are theory, averaging contrast decay in 1D chains initialized from a thermal distribution of the 3D cloud including  error bands stemming from $T_{2}$ uncertainties. \textbf{C}, Fitted contrast oscillation frequencies (black points) are compared to the calculated superexchange frequency (blue line) including bond-charge corrections to $t_{z}$, which averages the expected oscillations with local $\Delta E_j$ and $U$ along the imaging direction. Contrast oscillation frequencies are also fit to the function $\kappa \times \bar{J}_{SE}$ (red line),  with $\bar{J}_{SE}$ including no corrections to $t_{z}$, finding $\kappa = 1.42(7)$. Error bars are $1 \sigma$ (standard deviation) uncertainty of the fitted frequency $f$.
     \textbf{D},  Contrast curves approximately collapse when dark times are rescaled by the fitted oscillation frequency (red line) in Fig.~\subref{fig:fig4}{C}. A simple simulation sampling spin chains with different lengths and coupling strengths (gray dashed line) is overlaid.}
    \label{fig:fig4}
\end{figure*}

As the transverse confinement is increased further, only $s$-wave interactions remain relevant. When the system enters the strongly interacting regime doubly occupied lattice sites across the whole array are suppressed and coherent superexchange interactions dominate the quantum dynamics. In Figs.~\subref{fig:fig4}{A},~\subref{fig:fig4}{B} we show the contrast decay as a function of dark time for $V_{\perp} > V_{z}$, finding a clear oscillatory feature on timescales of the superexchange rate $\bar{J}_{\mathrm{SE}}$. For these measurements $V_{z}$ is fixed to  17.4$E_{R}$ at which $t_{z} \approx 14.2$ Hz. $\bar{J}_{\mathrm{SE}}$ is tuned by varying  $V_{\perp}$ between 19.7 and 67.4$E_{R}$, thus varying $U$ from 1.2 to 2.3 kHz. In the $V_{\perp}\gg V_z$  regime, the system is comprised of isolated vertical tubes along $z$ as shown in Fig.~\subref{fig:fig2}{C}. We assume all atoms within each tube are  pinned in place even for non-unit filling, since the local potential difference is much stronger than tunneling ($\Delta E_{j} \gg t_{z}$). We further assume that every uninterrupted chain of neighbouring atoms within a given tube undergoes evolution under the superexchange Hamiltonian $\hat{H}_{\mathrm{SE}}$. Their evolution is independent of other chains, and the contrast is an average over all chains. The curves in Figs.~\subref{fig:fig4}{A},~\subref{fig:fig4}{B} show numerical predictions averaging over the full 3D system using calibrated experimental parameters except entropy-per-particle in the lattice, which find good agreement when the overall slow decay in contrast reported in Fig.~\subref{fig:fig2}{C} is factored in.

\begin{figure*}[hbtp]
    \centering
    \includegraphics[width=5in]{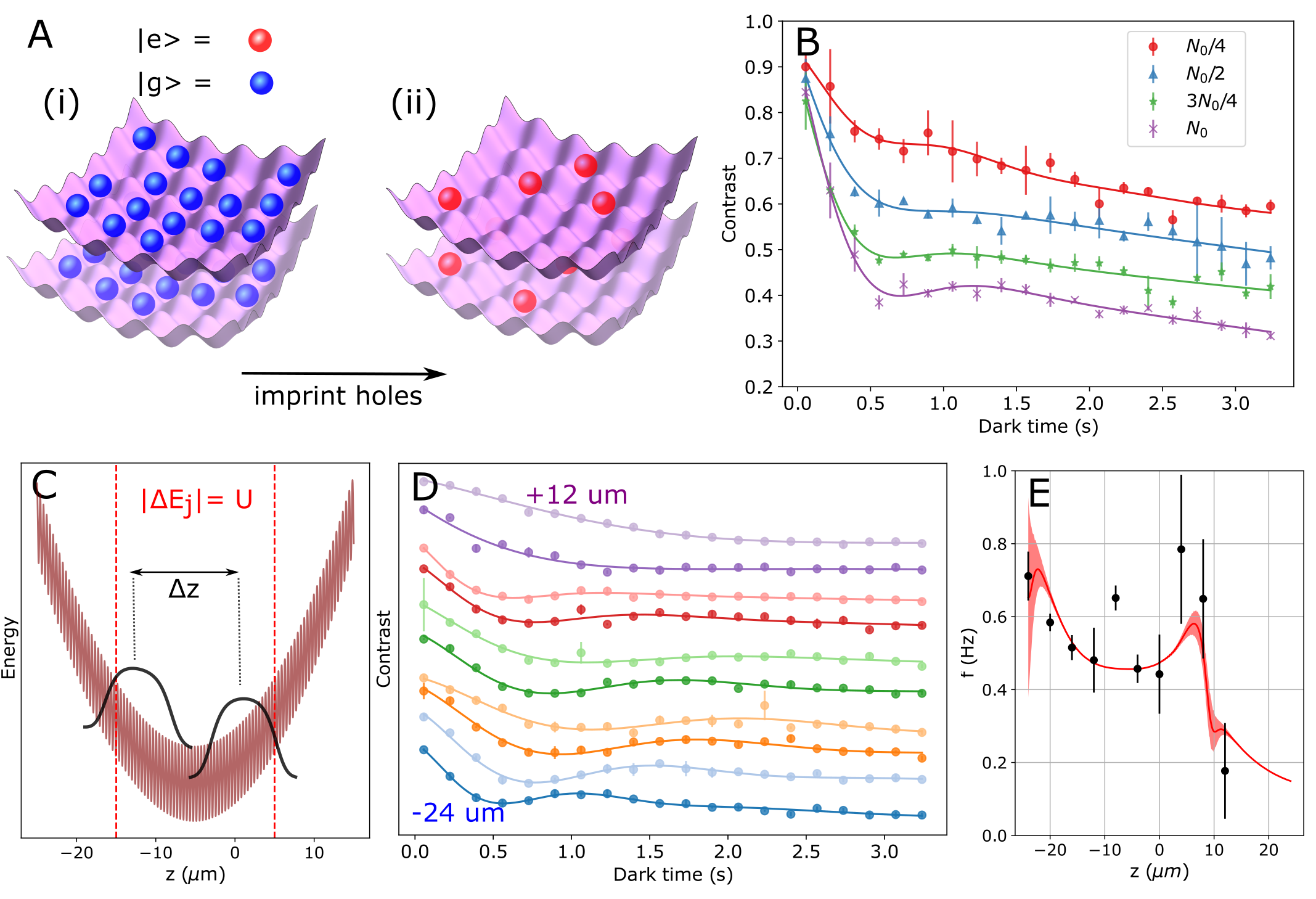}
    \caption{
	    \textbf{Controlling superexchange interactions}. All measurements presented here are performed at trap depths $V_z=17.4$ $E_{R}$ and $V_\perp=44.9$ $E_{R}$. The fraction of atoms participating in superexchange is modified by reducing the filling fraction via uniformly adding holes as depicted in panel \textbf{A}.  In (i), the initial state is a near unity filled sample of ground state atoms. Next, atoms are placed in a superposition state with tunable pulse area.  Light resonant with $\ket{^{1}S_{0}}$ is turned on to imprint holes, with the remaining atoms in $\ket{^{3}P_{0}}$ as shown in (ii). The contrast decay is plotted in \textbf{B} as the clock pulse area and thus total atom number $N$ is reduced compared to the initial atom number $N_{0}$. The solid lines shown in panels \textbf{B}, \textbf{D} are fits using the model $C_{\textrm{SE}}(T)$ provided in the main text. Error bars are $1 \sigma$ (standard deviation). In panel \textbf{C}, the superexchange coupling is modified by changing the position of the atoms in the lattice potential varying the site-to-site energy shift $\Delta E_{j}$. At the positions indicated by vertical red lines tunneling becomes resonant and strongly enhances the local $J_{\text{SE}}(j)$. However, averaged over the whole cloud this only slightly modifies the oscillation frequencies. Oscillations in contrast at different vertical positions $z$ are shown in panel \textbf{D}; curves are shifted vertically according to $z$ position. These measured oscillation frequencies are compared with a heuristic superexchange simulation (red line) of the Ramsey contrast in panel \textbf{E}~\cite{supplement}.}
    \label{fig:fig5}
\end{figure*}

To extract the measured superexchange rates, we vary $V_{\perp}$ and fit the experimentally measured contrast decay to the function $C_{\mathrm{SE}}(T) = A  e^{-T/T_{2}} \;+ \;B  \textrm{cos}(2 \pi f T)e^{-T/T_{\mathrm{osc}}}\; + \; D$. The measured oscillation frequencies $f$ are compared to a theoretically modelled superexchange frequency (blue line) accounting for the lattice inhomogeneity and bond-charge corrections of $t_{z}$ in Fig.~\subref{fig:fig4}{C}. To mitigate uncertainties of higher order corrections to Hubbard parameters like $t_{z}$, the contrast oscillation frequencies are also fit to the  function $\kappa \times \bar{J}_{\mathrm{SE}}$ (red line), where we determine $\kappa = 1.42(7)$. The simplified model for calculating $\bar{J}_{\mathrm{SE}}$ assigns individual $U$ and $\Delta E_j$ to localized atom pairs and averages the resulting local contrast oscillations in vertical direction to extract an oscillation frequency~\cite{supplement}. The agreement is good for all but the deepest $V_{\perp}$, for which the experimentally measured rate appears to be higher-frequency. Numerical calculations suggest this could arise from additional interaction inhomogeneity that favors higher frequency contributions. In Fig.~\subref{fig:fig4}{D}, the dark times of the contrast decay data are rescaled by the fitted superexchange rate $\kappa \times \bar{J}_{\mathrm{SE}}$ from Fig.~\subref{fig:fig4}{C}. The rescaled data collapse to a single curve, reflecting the underlying superexchange dynamics in all measurements. This is also in agreement with a theoretical model with randomly sampled spin chains of different lengths and coupling strengths to capture the effects of finite temperature and trap inhomogeneity without invoking explicit parameters (see discussion of Fig.~\subref{fig:fig5}{C - E}). We note that the lattice curvature is changing as a function of $V_\perp$, thus increasing $J_{\textrm{SE}}(j)$ inhomogeneity which prevents rescaling surpassing the measurement errors.




In order to study the properties of the interactions further, we vary the lattice filling and the energy offsets $\Delta E_{j}$ of the local lattice tilt in Fig.~\ref{fig:fig5}. First, the fraction of atoms participating in superexchange is reduced by imprinting holes in the lattice.  Beginning with maximum filling, before Ramsey spectroscopy a variable clock laser pulse duration is used to shelve atoms in $\ket{e}$ with spatially uniform probability, and subsequently the remaining $\ket{g}$ atoms are removed with resonant light at 461 nm (see Fig.~\subref{fig:fig5}{A}). The ensuing contrast decay as a function of the total atom number $N$ is plotted in Fig.~\subref{fig:fig5}{B}. The oscillation amplitude, reflecting the fraction of atoms participating in superexchange, is strongly decreased as $N$ is reduced due to the increasing number of holes. Due to the reduced filling fraction at the wings of the atom cloud, this effect is also observed when choosing the region of interest to be an annulus and increasing its radius compared to $P_{2}$\cite{supplement}.  


As the position of the atoms in the combined potential of gravity and the lattice confinement is shifted vertically the site-to-site energy shift $\Delta E_{j}$, and consequently the superexchange interaction strength, is strongly modified. We precisely move the cloud position at the $\mu$m scale~\cite{supplement}.  Figure~\subref{fig:fig5}{D} displays these oscillations as a function of cloud position $z$. We compare the oscillation frequency with a heuristic simulation analogous to  Fig.~\subref{fig:fig4}{C} of the Ramsey contrast in Fig.~\subref{fig:fig5}{E} (red line). Averaging the Ramsey signal along the $z$-direction during imaging strongly suppresses the effect of locally enhanced $J_{\mathrm{SE}} (j)$ where $U=\Delta E_j$. The asymmetry of the background trap gradient around $z=0$ leads to a reduction of the oscillation frequency at large $z$ where $\Delta E_j>U$. The frequency of the simulation shows qualitative agreement with the measured oscillation.

In conclusion, we have used our degenerate Fermi gas 3D optical lattice clock with anisotropic and tunable tunneling rates in the presence of spin-orbit coupling to directly probe different regimes of interaction effects described by the Fermi-Hubbard Hamiltonian. Superexchange interactions are identified as an important systematic effect that degrade the precision of optical lattice clocks operating with high filling at timescales $h/\bar{J}_{\textrm{SE}}$. We demonstrate that we can both microscopically model and control these interactions in the 3D optical lattice. 

For clock metrology, we can either reduce the magnitude or control the form of the superexchange interactions to enhance clock performance.  For example, we can increase the lattice constant $a$ sufficiently large to reduce the tunneling rate to a negligible value~\cite{hutson2019engineering}.
Alternatively, a variable lattice spacing can be used to make $a$ commensurate with $\lambda_{clk}$ to achieve $\varphi$ mod 2$\pi$ = 0.
Without  SOC ($\varphi$ mod 2$\pi$ = 0) the isotropic Heisenberg Hamiltonian $  \sum \limits_{j} J_{\mathrm{SE}}(j) \hat{\bold{s}}_{j} \cdot \hat{\bold{s}}_{j + 1}$  is recovered, and any coherent spin state becomes an eigenstate accumulating only a trivial global phase. 
On the other hand,  collective superexchange interactions can be used to produce spin entanglement for quantum enhanced sensing~\cite{perlin2020spin}. At intermediate in-plane tunnelling rates, these isotropic, Heisenberg interactions couple the single particles within each plane to collective spins~\cite{mamaev2023spin}. Thus, by reducing single-particle inhomogeneities via potential shaping or layer selection \cite{gaunt2013bose}, the collective spins across all planes can be squeezed by SOC-induced XXZ interactions investigated here.

\textbf{Acknowledgement} We thank A. Aeppli, Z. Hu, J. Hur, D. Kedar, K. Kim, M. Miklos, J. M. Robinson, Y. M. Tso, W. Warfield, Z. Yao for useful discussions. We thank A. M. Kaufman  and N. D. Oppong for careful reading of the manuscript and for providing insightful comments. Funding for this work is provided primarily by DOE Center of Quantum System Accelerator and also by NSF QLCI OMA-2016244, NSF JILA-PFC PHY-2317149, V. Bush Fellowship, AFOSR FA9550-18-1-0319,  and NIST.

\textbf{Author contributions}

The experiment was performed by W.R.M., S.L., L.Y., B.L., M.N.F., R.B.H, and J.Y. The theory model was developed by M.M, A.C., and A.M.R. All authors contributed to data analysis and writing of the manuscript. 

\textbf{Data availability}

All data is available from the corresponding
authors on reasonable request.

\textbf{Competing interests}

The authors declare no competing interests.

\newpage

\clearpage
\section*{Supplementary Materials}

\renewcommand{\thefigure}{S\arabic{figure}}
\setcounter{figure}{0}

\subsection*{State preparation}

The experimental sequence used on our three-dimensional optical lattice clock is detailed in previous publications from this experiment \cite{sonderhouse2020thermodynamics, milner2023high, hutson2019engineering}. In this section, we briefly summarize our procedure to prepare a nuclear spin-polarized degenerate Fermi gas. This quantum gas is the starting point before lattice loading and clock spectroscopy. First, approximately $100 \times 10^{6}$ atoms are trapped at $\approx$ mK temperatures in our "blue" MOT using the $\Gamma = 2 \pi \times 30.5$  MHz, $^{1}S_{0}$ $\leftrightarrow$ $^{1}P_{1}$ transition.  Next, our "red" MOT is prepared on the $^{1}S_{0}$ $\leftrightarrow$  $^{3}P_{1}$ transition with linewidth $\Gamma = 2 \pi \times 7.5$ kHz. Given the comparatively smaller $^{3}P_{1}$ transition linewidth and thus Doppler temperature,  $10 \times 10^{6}$ atoms are laser cooled to $1 \mu$K temperatures in a 1064 nm crossed dipole trap (XODT). With approximately an equal distribution of all 10 nuclear spin states $\ket{^{1}S_{0}, m_{F} = -9/2,  m_{F} = -7/2 ... m_{F} = 9/2}$, forced evaporation is employed to cool the atoms to a degeneracy parameter of $T/T_{F} \approx 0.1$ with $30 \times 10^{4}$ atoms per spin state \cite{sonderhouse2020thermodynamics}.  For clock spectroscopy,  a single Zeeman sublevel is desired, requiring nuclear-spin polarization. For polarization, we apply a spin dependent potential detuned from the $^{3}P_{1}$ transition where the polarizability is negative for all sub-levels except $|m_{F} = 9/2|$. Including an initial optical pumping stage, we prepare a nuclear spin polarized gas in $\ket{^{1}S_{0}, m_{F} = -9/2}$ with temperature $T/T_{F} \approx 0.2$ \cite{milner2023high}. We note that the lattice lifetime depicted in Fig.~S1 is 108(5) s, much longer than the Ramsey dark times probed in this work. 

\begin{figure*}[hbtp]
    \centering
    \includegraphics[width=3in]{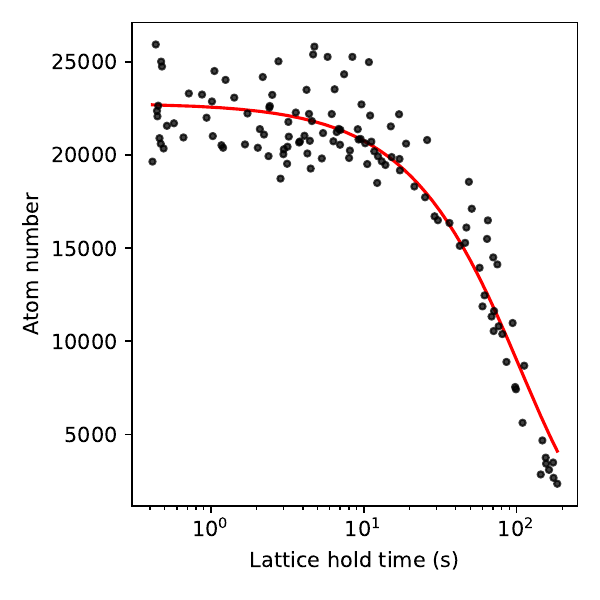}
    \caption{
	    \textbf{$^{1}S_{0}$ lattice lifetime}. The atoms are held in a deep 3D lattice and the atom number loss is measured as a function of hold time. A 1/$e$ time of 108(5) seconds is fit to the data. The atom loss is likely limited by a combination of parametric heating from trap and vacuum lifetime. This technical loss timescale is long compared to all dynamics studied in this work. }
    \label{fig:figs1}
\end{figure*}

\subsection*{Lattice potential}

To confine the atoms in our 3D lattice, we adiabatically ramp all lattice beams in three 150 ms steps. Starting at 0 $E_{R}$, we ramp to 2.5 $E_{R}$, then 10 $E_{R}$, before ramping to our final trap depth $V_{F}$ for clock spectroscopy. For measurements with $V_{F} \leq$ 10 Er, we ramp $ 0 E_{R} \rightarrow 2.5 E_{R} \rightarrow V_{F}$ then hold. To prepare for clock spectroscopy in our magic wavelength lattice, we ramp off the XODT trap over 100 ms, while leaving the lattice depths at $V_{F}$.

The superexchange coupling strength $J_{\mathrm{SE}}(j)$ depends on the site-to-site energy shift $\Delta E_{j}$ between atoms on lattice sites $j$ along the clock axis $z$. Calculating $\Delta E_{j}$ requires determining the atom's location in the lattice Gaussian confinement during clock spectroscopy. We go step-by-step through the loading procedure, considering the trapping potential at each step, to arrive at the final confinement. 

First, we consider the trapping potential in the XODT before lattice loading. The XODT potential including the gravitational tilt is $U_{XODT}(j) / \hbar$ = ($1/2$ $m\omega_{XODT}^2a^2) (j-j_{\textrm{XODT}})^2 + mga j$, where $j$ is a dimensionless position in units of the lattice spacing, $j_{\textrm{XODT}}$ is the center position of the XODT beam, and $\omega_{XODT}$ = $2 \pi \times 250$ Hz, thus $U_{XODT}(j) / \hbar$ = ($44.7$ Hz) $(j-j_{\textrm{XODT}})^2$ + ($873.1$ Hz) $j$. The bottom of this displaced parabola is at $j_{\textrm{s}}$, where the atomic density distribution is centered.

Next, we consider the potential of the optical lattice. We align the center of the lattice beams to the atom cloud in the XODT. Thus, the center of the transverse lattice beams is located at $j_{\textrm{s}}$. After ramping down the XODT beams, the offset of the vertical lattice potential, generated by the two transverse lattice beams, is therefore $U_{\textrm{lat}}=-2V_\perp\exp(-2a^2(j-j_{\textrm{s}})^2/w^2) + mgaj$, where $w$ is the transverse lattice beam waist radius extracted in Fig.~S3. Thus, the minimum of $U_{\textrm{lat}}$ is at $j<j_{\textrm{s}}$. Because we turn off the 1064 nm trap in a deep lattice where tunneling is frozen, the atom position is centered at $j_{\textrm{s}}$ throughout the Ramsey spectroscopy. Therefore, the cloud is significantly displaced from the center of the lattice potential, as depicted in Fig.~\subref{fig:fig5}{C}. At shallow transverse lattice depths where the Gaussian confinement is weak, Wannier-Stark peaks at approximately $mga$ are observed as seen in Fig.~S2. As the transverse confinement, and thus the inhomogeneity, is increased these peaks are broadened. 

\begin{figure*}[hbtp]
    \centering
    \includegraphics[width=5in]{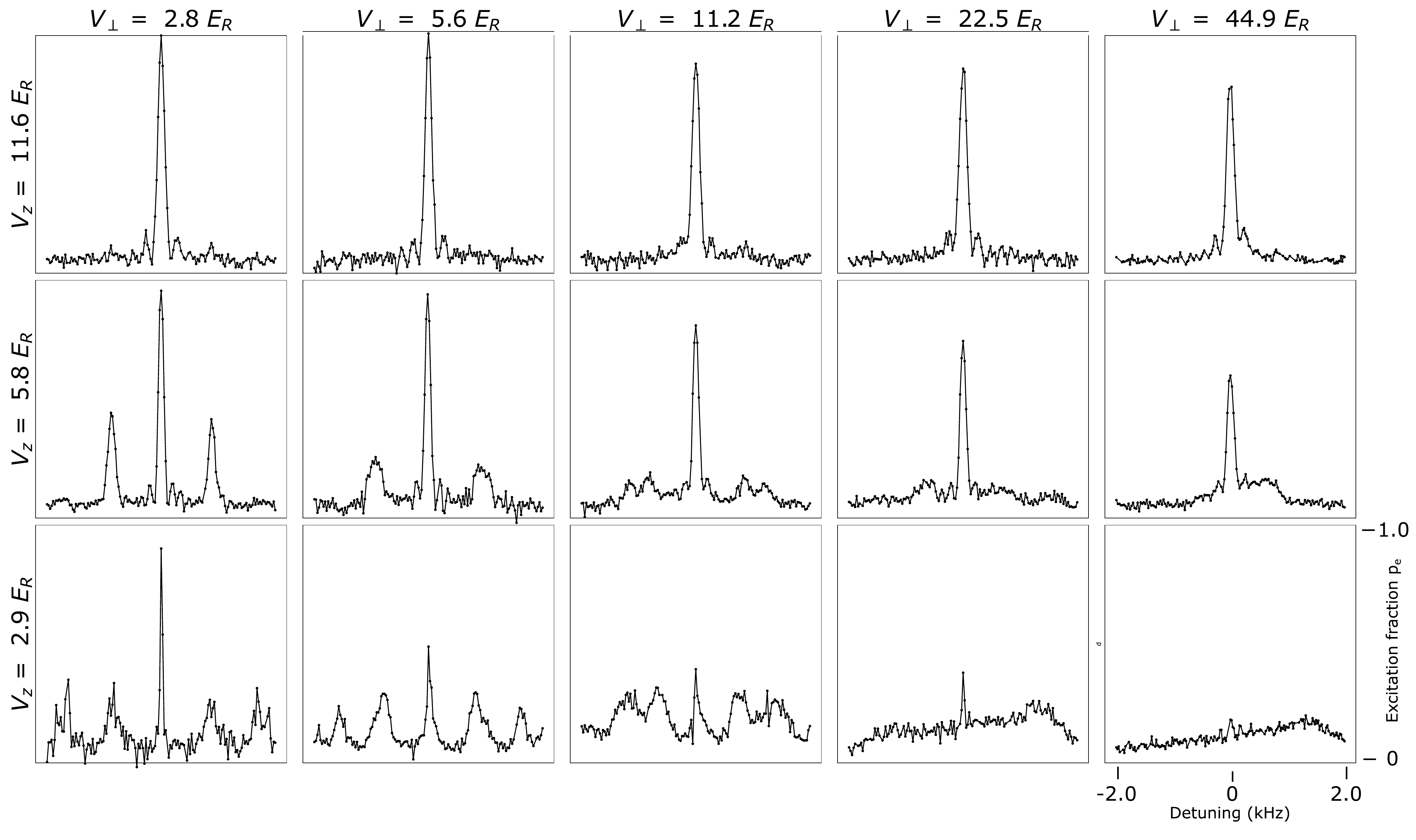}
    \caption{
	    \textbf{Shallow lattice Rabi spectroscopy}. Rabi spectroscopy in shallow lattice confinement is studied. The detuning span in each measurement is -$2 \mathrm{kHz}$ to $2 \mathrm{kHz}$. The Rabi pulse duration was optimized for each measurement to maximize the excited state fraction at zero detuning. }
    \label{fig:figs2}
\end{figure*}


The XODT trap frequency $\nu_{XODT}$ is calibrated via sloshing measurements, where the atoms are displaced in the dipole trap and their spatial oscillations at $\nu_{XODT}$ are measured. We do a similar measurement to determine the Gaussian confinement from the lattice beams. Here, the atoms are loaded from the XODT into a 2D lattice formed by both transverse lattice beams. As depicted in Fig.~S3, the oscillation frequency is plotted as a function of transverse trap depth $V_{\perp}$.

\begin{figure*}[hbtp]
    \centering
    \includegraphics[width=4.5in]{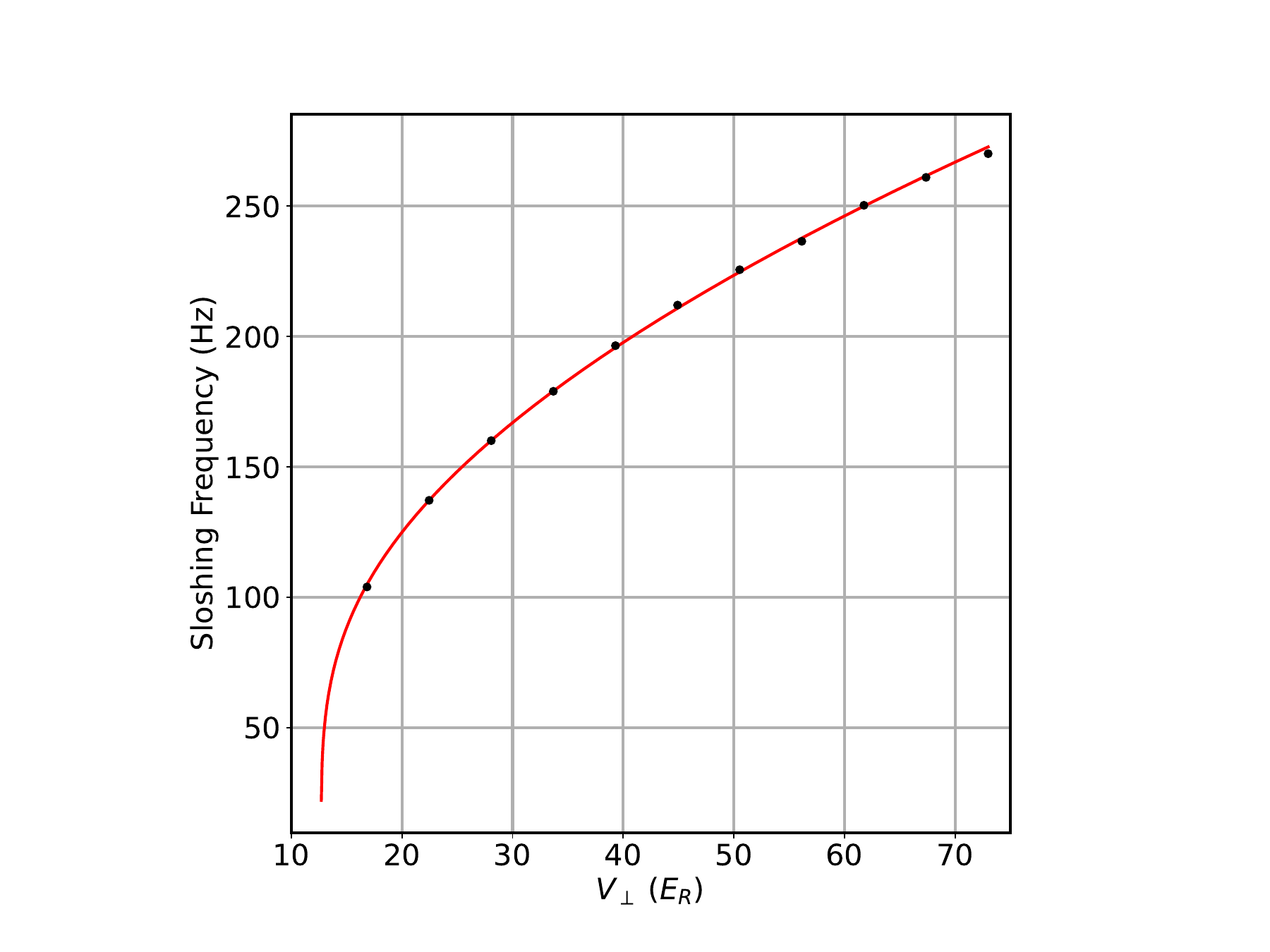}
    \caption{
	    \textbf{Lattice Gaussian confinement calibration}. The trap frequency is measured for atoms solely confined in both horizontal lattice beams, each at a power equivalent to the trap depth $V_\perp$. The oscillations are initiated by rapidly switching off the superimposed horizontal dipole trap. We note, that for the evaluation the gravitational sag needs to be taken into account, and that employing a harmonic approximation for estimating the sag induces discernible errors in the beam parameters. Numerical evaluation of the Gaussian beam curvature at the position of the atomic sag allows the construction of a fit function from which the beam waist radius is extracted to be $w=\SI{62.2(14)}{\micro\meter}$.}
    \label{fig:figs3}
\end{figure*}

 For the measurements in Fig.~\subref{fig:fig5}{D}, we displace the position of the atoms in the lattice potential to locally change $\Delta E_{j}$. This requires precise adjustment of the density distribution at the \si{\micro\meter} level. To accomplish this, we use our horizontal imaging system with imaging parameters in \cite{milner2023high}. We used a commercial, piezo actuated mirror that moves the XODT beam to adjust the center position of the atomic density distribution.


\subsection*{Ellipse fitting analysis}

Our saturated imaging procedure using our high-resolution (NA = 0.2) vertical imaging system is detailed in \cite{milner2023high}. We extract the contrast of our atomic ensemble using imaging spectroscopy where the local excitation fractions in two spatially separated regions are compared. Ellipse fitting is used, where parametric plots in regions $P_{1}$, $P_{2}$ trace an ellipse $= \frac{1}{2} + \frac{C}{2}\mathrm{cos}\Bigl(2 \pi f_{1, 2} T + \phi_{0}\Bigr)$  \cite{marti2018imaging}. The differential frequency shift $f_{1, 2}$ between spatial regions of interest following our XY8 decoupling sequence is consistent with zero. This is illustrated in Fig.~\subref{fig:fig2}{B}, where the parametric plots show a straight line with no opening angle. In regions-of-interest much larger than $P_{2}$, we observe that the fraction of atoms participating in superexchange is reduced due to the reduced filling fraction in the thermal wings of the cloud as shown in Fig.~S4. We use jackknifing to determine the $1 \sigma$ (standard deviation) contrast error bars. For $n$ excitation fraction measurements at a given dark time $T$, we first determine the contrast $\bar{C}$ using all data points. Next, we cycle through all $n$ datapoints and recompute the contrast  $C_{\neq i }$ excluding the $i$-th data point. Summing all contributions we estimate the uncertainty: Var($C$) = $\frac{n - 1}{n} \sum \limits_{i}^{n} (\bar{C} - C_{\neq i })^{2}$ 

\begin{figure*}[hbtp]
    \centering
    \includegraphics[width=5in]{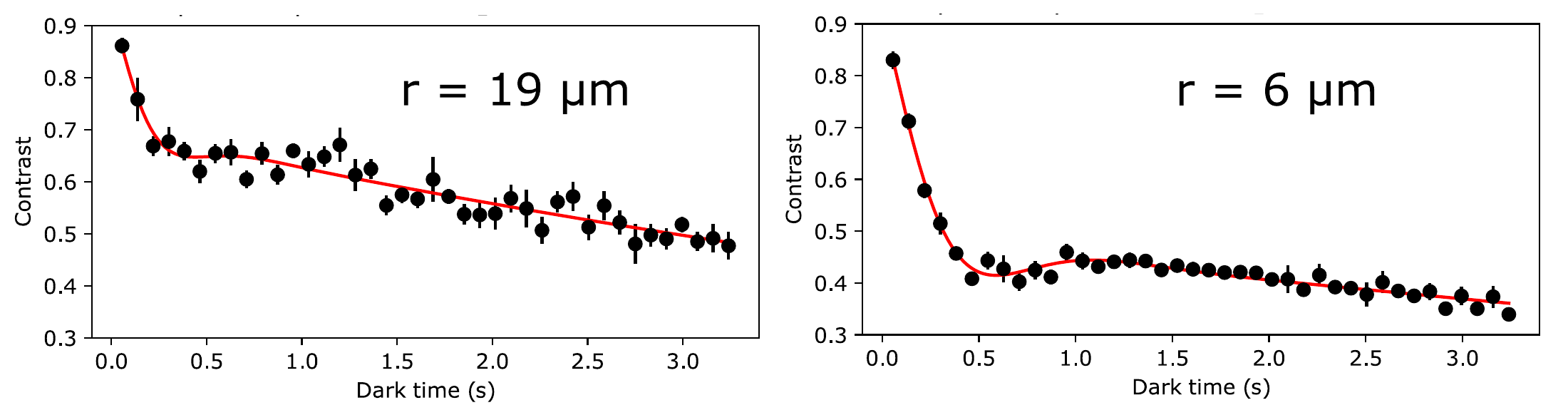}
    \caption{
	    \textbf{Radial dependence on contrast}. Contrast decay in two regions of interest using annulus' with thickness \SI{2}{px} (\SI{0.8}{\micro\meter}) and radii \SI{14}{px} (\SI{6}{\micro\meter}),  \SI{48}{px} (\SI{19}{\micro\meter}) are plotted. $V_{\perp}$ = 44.9 $E_{R}$ and $V_{Z}$ = 17.4 $E_{R}$ was used for these measurements. Error bars are $1 \sigma$ (standard deviation) uncertainty of the Ramsey contrast obtained from jackknifing. The sparsely filled region (\SI{19}{\micro\meter}) has higher contrast due to the lower filling fraction compared to near the center of the cloud (\SI{6}{\micro\meter}). }
    \label{fig:figs4}
\end{figure*}

\subsection*{Two-well Fermi-Hubbard Hamiltonian}
To understand the superexchange dynamics, we start from the two-well Fermi-Hubbard Hamiltonian with a energy tilt $\Delta E$ between sites,
\begin{equation} \label{eq:fermi-hubbard}
    \hat{H}_{\mathrm{FH}}=-t_z\sum_{\sigma\in \{g,e\}}(\hat{\tilde{c}}^{\dag}_{0,\sigma}\hat{\tilde{c}}_{1,\sigma}+H.c.)+U\sum_{j\in \{0,1\}}\hat{n}_{j,e}\hat{n}_{j,g}+\frac{\Delta E}{2}(\hat{n}_{1}-\hat{n}_{0}),
\end{equation}
Here, $\hat{\tilde{c}}_{j, \sigma}^{\dag} (\hat{\tilde{c}}_{j, \sigma})$ creates (annihilates) a fermion on site $j$ with spin $\sigma$ in the lab frame. We define $\hat{n}_{j,\sigma}=\hat{\tilde{c}}_{j, \sigma}^{\dag}\hat{\tilde{c}}_{j, \sigma}$, and $\hat{n}_j=\hat{n}_{j,e}+\hat{n}_{j,g}$. In the Mott-insulating limit $U\gg t_z$, the double-occupied states are separated by a large energy gap $\sim U$, which allows for restriction of dynamics in the single-occupied states via second-order perturbation theory. We get the effective Hamiltonian for superexchange interaction,
\begin{equation}
    \hat{H}_{\mathrm{eff}}=J_{\mathrm{SE}}\bigg(\hat{\tilde{\mathbf{s}}}_0\cdot\hat{\tilde{\mathbf{s}}}_1-\frac{1}{4}\bigg),
    \label{eq:eff}
\end{equation}
where
\begin{equation} \label{eq:methods_jse}
    J_{\mathrm{SE}}=\frac{4t_z^2U}{U^2-\Delta E^2}.
\end{equation}
Here the spin operators are defined as $\hat{\tilde{\mathbf{s}}}_{j}=\sum_{\alpha\beta=\{e,g\}}\hat{\tilde{c}}^{\dag}_{j,\alpha}\bm{\sigma}_{\alpha\beta}\hat{\tilde{c}}_{j,\beta}/2$, where $\bm{\sigma}_{\alpha\beta}$ are Pauli matrices. 
Expanding $\hat{\tilde{\mathbf{s}}}_0\cdot\hat{\tilde{\mathbf{s}}}_1=(\hat{\tilde{\mathbf{s}}}_0+\hat{\tilde{\mathbf{s}}}_1)^2/2-3/4$, we obtain triplet states ($|d\rangle\equiv|g,g\rangle$, $|u\rangle\equiv|e,e\rangle$,$|t\rangle\equiv(|g,e\rangle+|e,g\rangle)/\sqrt{2}$) with zero  energy, and the singlet state ($|s\rangle\equiv(|g,e\rangle-|e,g\rangle)/\sqrt{2}$) with energy $-J_{\mathrm{SE}}$ with respect to the triplet states.

The oscillation of the Ramsey contrast can be understood based on the time evolution of the initial state generated by the first Ramsey pulse. This state has a spiral phase $e^{i j \varphi}$ imprinted by the clock laser with Rabi frequency $\Omega$, $\hat{H}_{\mathrm{clock}}(\theta)/\hbar=\frac{1}{2}\sum_j(|\Omega|e^{i\theta}\times \hat{\tilde{s}}^{+}_je^{ij\varphi}+H.c.)$, where $\theta$ controls the rotation axis. We consider the initial state generated by the first Ramsey pulse with $|\Omega| T_1=\pi/2$,
\begin{equation}
\begin{aligned}
\ket{\psi_{\textrm{init}}} &= e^{-i\hat{H}_{\mathrm{clock}}(\theta=\pi/2)T_1/\hbar}\ket{g}_0\otimes\ket{g}_1\\
&= \frac{1}{\sqrt{2}}(\ket{g}_0 + \ket{e}_0) \otimes \frac{1}{\sqrt{2}}(e^{-i \varphi/2} \ket{g}_1 + e^{i \varphi/2}\ket{e}_1)\\
&= \frac{1}{2} \left[e^{-i \varphi/2}\ket{d} + e^{i \varphi/2}\ket{u}+\sqrt{2}\cos(\varphi/2)\ket{t} + i \sqrt{2}\sin(\varphi/2) \ket{s}\right].
\end{aligned}
\end{equation}
For simplicity, we ignore all the echo pulses during the dark time. So the dynamics in the dark time can be described by the singlet state $|s\rangle$ acquiring a phase $e^{i J_{\mathrm{SE}} T/\hbar}$. Then we apply the second Ramsey pulse with $|\Omega| T_2=\pi/2$ with the same $\theta$, and get the final state 
\begin{equation}
    \begin{aligned}
    \ket{\psi_f}&=-ie^{i\varphi/2}\sin(J_{\mathrm{SE}}T/2\hbar)\sin^2(\varphi/2)\ket{d}+\frac{1}{4}e^{-i\varphi/2}\big(3e^{-iJ_{\mathrm{SE}}T/2\hbar}+(1-\cos(\varphi))e^{iJ_{\mathrm{SE}}T/2\hbar}+\cos(\varphi)\big)\ket{u}\\
    & + \frac{i}{\sqrt{2}}\sin(J_{\mathrm{SE}}T/2\hbar)\sin(\varphi/2)\sin(\varphi)\ket{t} - \frac{1}{\sqrt{2}}\sin(J_{\mathrm{SE}}T/2\hbar)\cos(\varphi/2)\sin(\varphi)\ket{s}.
    \end{aligned}
\end{equation}

In this case the Ramsey contrast is given by $C=2|\langle \psi_f|\hat{\tilde{s}}^z|\psi_f\rangle|/N$, where $\hat{\tilde{s}}^z= \hat{\tilde{s}}^z_1+\hat{\tilde{s}}^z_2$. The Ramsey contrast will thus undergo oscillatory dynamics, 
\begin{equation}
    C(T)=\bigg|\cos^2\bigg(\frac{\varphi}{2}\bigg)+\sin^2\bigg(\frac{\varphi}{2}\bigg)\cos(J_{\mathrm{SE}} T/\hbar)\bigg|.
    \label{eq:contrast}
\end{equation}

\subsection*{Heuristic averaging of superexchange dynamics}
In our experiment operating under the conditions $V_z\ll V_{\perp}$, we ignore tunneling in the transverse directions and consider superexchange dynamics only along the $z$ direction.
To capture the superexchange dynamics, we need to include a spatially varying superexchange rate $J_{\mathrm{SE}}(j)$, due to the tilt generated by gravity and the confinement generated by the Gaussian profile of the lattice beam. In addition to the site-to-site energy shift $\Delta E_j$, the reduction of the transverse lattice power at $\lvert j-j_{\textrm{s}}\rvert\gtrsim w/a$ (with lattice constant $a$) also decreases the on-site interactions $U_j$ and therefore induces a weak $j$-dependence. Instead of applying Eq.~(\ref{eq:methods_jse}), to avoid artifacts from the divergence present in this approximation, we obtain $J_{\mathrm{SE}}(j)$ from an independent diagonalization of Eq.~(\ref{eq:fermi-hubbard}) at each lattice site $j$. For simplicity, we focus on the region with uniform density in the $x$-$y$ plane and therefore consider the variation of $J_{\textrm{SE}}$ only in vertical direction.

Here we assume the oscillatory dynamics are mainly generated by two-atom chains. So we can heuristically generalize Eq.~(\ref{eq:contrast}) to average over all possible local one-atom and two-atom chains in our system. We define $n(j)$ as the local filling fraction of lattice sites labelled by $j$, such that the total atom number in each vertical tube is $N_{\textrm{tube}}=\sum_jn(j)$. In the following, two adjacent lattice sites along $z$ direction are considered but all quantities are given with respect to individual lattice sites. The probability to have only one atom in these two lattice sites is $p^{(1)}(j)=n(j)[1-n(j)]$, and the probability to have two atoms is $p^{(2)}(j)=[n(j)]^2$. Here, the magnetization of a one-atom chain is $\langle\tilde{s}^{z,(1)}(j)\rangle=\langle \psi_f|\hat{\tilde{s}}^z|\psi_f\rangle=1/2$, and in a local two-atom chain $\langle\tilde{s}^{z,(2)}(j)\rangle=\{\cos^2(\varphi/2)+\sin^2(\varphi/2)\cos[J_{\mathrm{SE}}(j) T/\hbar]\}/2$ (see Eq.~(\ref{eq:contrast})) and thus the average magnetization per size is $\langle\tilde{s}^z(j)\rangle=\sum_kp^{(k)}\langle\tilde{s}^{z,(k)}\rangle$. We then average over all spatial positions to obtain the Ramsey contrast
\begin{align} \label{eq:avg}
  \begin{split}
    C(T)&=\frac{1}{N_{\textrm{tube}}}\bigg|\sum_j2\langle\tilde{s}^z(j)\rangle\bigg|\\
    &=\bigg|1-\sum_j\frac{[n(j)]^2}{N_{\textrm{tube}}}\sin^2\bigg(\frac{\varphi}{2}\bigg)+\sum_j\frac{[n(j)]^2}{N_{\textrm{tube}}}\sin^2\bigg(\frac{\varphi}{2}\bigg)\cos(J_{\mathrm{SE}}(j) T/\hbar)\bigg|.
  \end{split}
\end{align}
By fitting a function of the from $Ae^{-T/\tau_{\textrm{osc}}}\cos(\bar{J}_{\mathrm{SE}}T/\hbar)$ to the oscillatory term in $C(T)$ we obtain the oscillation frequency $\bar{J}_{\mathrm{SE}}/h$, which is the basis for the solid red lines in Figs.~\subref{fig:fig4}{C} and \subref{fig:fig5}{E}. The bond-charge corrections $\Delta t_z$ taken into account for the blue line in Fig.~\subref{fig:fig4}{C} are scaled with $U_j$ to account for the inhomogeneity across the atom cloud. The error bands are derived from the uncertainty of this fit parameter. We choose a temperature of $\sim\SI{370}{nK}$ to roughly match the peak-to-peak oscillation amplitude of $2\sin^2(\varphi/2)\sum_j[n(j)]^2/N_{\textrm{tube}}\sim0.7$ observed in most measurements (cf.~Figs.~\subref{fig:fig5}{D} or \subref{fig:fig4}{B}).

We note that this approach of estimating the oscillation frequency does not provide a comprehensive and quantitative model for the coherence as it neglects effects from longer chains and the dynamical decoupling pulse sequence. These are taken into account in the following section.

\subsection*{Superexchange contrast dynamics}
Now we generalize Eq.~(\ref{eq:eff}) to the case of many-atoms. Here we  work in a ``spiral'' frame where the initial state is uniform (all atoms in the same superposition state) and the site-dependent laser phase $\varphi$ is absorbed into the spin operators across the lattice, $\hat{s}^{\pm}_j=\hat{\tilde{s}}^{\pm}_je^{\pm ij\varphi}$, $\hat{s}^{Z}_j=\hat{\tilde{s}}^{Z}_j$, where $j$ is the lattice index along $z$ direction. 
This transformation lead to the following 1D spin-spin interaction Hamiltonian~\cite{mamaev2021tunable},
\begin{equation}
\begin{aligned}
\hat{H}_{\mathrm{SE}} &= \sum_{j=1}^{L-1} J_{\mathrm{SE}}(j) \left[\cos(\varphi) \left(\hat{s}_{j}^{X}\hat{s}_{j+1}^{X} + \hat{s}_{j}^{Y}\hat{s}_{j+1}^{Y}\right) + \hat{s}_{j}^{Z}\hat{s}_{j+1}^{Z} + \sin(\varphi) \left(\hat{s}_{j}^{X}\hat{s}_{j+1}^{Y} - \hat{s}_{j}^{Y}\hat{s}_{j+1}^{X}\right)\right],\\
J_{\mathrm{SE}}(j) &= \frac{4t_{z}^2 U}{U^2 - \Delta E_{j}^2},
\end{aligned}
\end{equation}
where $L$ is the number of sites, and $J_{\mathrm{SE}}(j)$ the superexchange interaction strength. The latter depends on both the on-site Hubbard interactions and the local potential difference $\Delta E_{j}=\frac{1}{2}m \omega_{lat}^2 a^2[(j+1-j_0)^2 - (j-j_0)^2]$, where $j_0$ is the bottom of the lattice confining potential.

The first three terms act as an XXZ Hamiltonian with spin anisotropy $\sim \sec(\varphi)$, which induces contrast decay. The last two terms are a Dzyaloshinskii-Moriya (DM) type interaction, which breaks exchange symmetry due to the chirality of the imprinted clock laser phase. The latter has been studied in the context of exotic chiral properties such as skyrmions. At the collective mean-field level such an interaction has no effect. In our case since the interaction strengths $J_{\mathrm{SE}}(j)$ are inhomogeneous, the DM interaction will also generate contrast decay, as each atom will feel an unequal force from its left and right neighbours due to the lack of exchange symmetry.

The Ramsey decay dynamics are modeled by initializing a product state of all spins in a uniform superposition state following the first Ramsey pulse as written above,
\begin{equation}
\ket{\psi_{\textrm{init}}} = e^{-i \frac{\pi}{2}\sum_{j} \hat{s}_{j}^{Y}}\bigotimes_{j} \ket{\downarrow}_{j}.
\end{equation}
The chain then undergoes time-evolution under the Hamiltonian, interspersed with echo pulses during the XY8 sequence. For a sequence including a single echo pulse we write,
\begin{equation}
\ket{\psi_f (t)} = e^{-i \hat{H}_{\mathrm{SE}} t/2} e^{-i \pi \sum_{j} \hat{s}_{j}^{X}} e^{-i \hat{H}_{\mathrm{SE}} t/2} \ket{\psi_{\textrm{init}}}.
\end{equation}
An XY8 sequence instead applies eight pulses about different axes as depicted in the main text Fig.~\subref{fig:fig2}{A}. After time-evolving the state, a final Ramsey pulse with an arbitrary phase $\theta$ is performed,
\begin{equation}
\ket{\psi_{f,\theta}(t)} = e^{-i \frac{\pi}{2}\sum_{j}\left[\cos(\theta) \hat{s}_{j}^{Y} + \sin(\theta)\hat{s}_{j}^{X}\right]}\ket{\psi_f(t)}.
\end{equation}
The contrast of the uninterrupted chain is obtained by measuring the excited state fraction,
\begin{equation}
N_{\theta}(t) = \bra{\psi_{f,\theta}(t)}\left[\sum_{j}\left(\hat{s}_{j}^{Z} + \frac{1}{2}\right)\right] \ket{\psi_{f,\theta}(t)}.
\end{equation}
For a single independent chain, contrast is obtained via,
\begin{equation}
C(t) = \frac{1}{L}\left[\text{max}_{\theta} N_{\theta}(t) - \text{min}_{\theta} N_{\theta}(t)\right].
\end{equation}
If there are multiple independent chains, their contributions to the excited state fraction $N_{\theta}(t)$ must be summed together for each angle $\theta$ before performing the maximization and minimization above.

A single site $L=1$ has unity contrast $C=1$ at all times. Chains with few sites will exhibit persistent oscillations of contrast, whereas chains with many sites will undergo decay, with revivals only occuring on timescales $\sim 1/L$. Inhomogeneity in the superexchange couplings $J_{\mathrm{SE}}(j)$ will also wash out revivals or oscillatory dynamics at longer times. The contrast dynamics from many summed, disordered chains thus generally exhibits only one or two oscillations before saturating to a constant value determined by how many of the chains had isolated single sites.

We compare this theory to the experiment by using specific lattice parameters, interaction coupling coefficients, and averaging over the 3D distribution, as detailed in the \textit{Full simulation of the 3D cloud} section of the Supplementary Materials, which yields a theoretically predicted contrast $C(T)$. Since the experiment also finds a slower decay on timescale $\sim 1/T_2$ measured in Fig.~\subref{fig:fig2}{C}, we normalize the resulting contrast obtained from numerical simulation of the superexchange Hamiltonian by a further factor $C(T)\to C(T) e^{-T/T_2}$. The resulting theoretically predicted contrast is shown as solid lines in Figs.~\subref{fig:fig4}{A},~\subref{fig:fig4}{B} of the main text, which is in good agreement with the measurements. The shaded region on the theory curves corresponds to an uncertainty of $\pm 2$ s for the $T_2$ in this adjustment factor (in line with the $T_2$ measurement uncertainty).

In addition, we provide a more simple theoretical prediction without invoking explicit experimental conditions. We randomly sample a large number of chains with lengths $L$ drawn from a Poisson distribution $P(\lambda)$ with low Poisson parameter $\lambda < 1$, appropriate for an initial thermal distribution. The coupling strengths $V_{j}$ in each chain are drawn from a Gaussian distribution of mean $J_{\mathrm{SE}} (j)$ and standard deviation $\epsilon J_{\mathrm{SE}} (j)$, with $\epsilon$ meant to capture inhomogeneity in the superexchange interactions. As $\epsilon$ increases, the contrast oscillations reduce in amplitude to the profile observed in the experiment. The curve in Fig.~\subref{fig:fig4}{D} of the main text shows the prediction for Poisson parameter $\lambda=0.25$ and $\epsilon = 0.4$. This curve is also adjusted by a factor of $e^{-\frac{1}{5}T \overline{J}_{\mathrm{SE}}/h}$ to account for slower atomic decay, using an effective lifetime of five superexchange cycles, which is in line with the experimental lifetimes and yields good agreement with all measured data.

\subsection*{Corrections to Fermi-Hubbard Parameters}
Due to wavefunction overlap with adjacent sites, additional interaction and tunneling terms are present in the Fermi-Hubbard Hamiltonian.
We identify the main contributions to be bond-charge type effects \cite{lühmann2012multiorbital} and an admixture of higher bands due to the gravitational tilt.

Bond-charge interactions are those with interactions between adjacent sites and additionally an exchange of the particles. This can be cast into the form of a tunneling term, thus directly correcting the tunneling energy $t_z'=t_z+\Delta t_z$ with
\begin{equation}
  \Delta t_z=-\frac{4\pi\hbar^2a_{eg}^-}{m}\int d^3x\,\psi_0^3\psi_1,
\end{equation}
where $\psi_j=\psi(x,y,z-ja)$ describes the ground band Wannier function $\psi$ at lattice site $j$. For lattice depths of $V_z=17.4$ $E_R$ and $V_\perp=44.9$ $E_R$ we obtain $\Delta t\approx h\times\SI{1.2}{Hz}$, which corresponds to an increase of about 8\% with respect to the bare value of $t_z\approx h\times\SI{14.2}{Hz}$.

A direct calculation of the Wannier-Stark wavefunction suggests an additional correction to the tunneling energy on the order of $\sim10\%$.
However, the exact calculation of the full contribution remains challenging because we estimate that all higher bands would be needed to be taken into account for a faithful quantification \cite{goban2018emergence}. Because these effects are barely above our experimental uncertainty we are mostly neglecting these corrections in this work.

\subsection*{1D large-spin Hamiltonian}
To model the contrast decay in Fig.~\ref{fig:fig3} we describe the $V_z\gg V_{\perp}$ regime in 3D optical lattice clocks using the assumption that the spins in each pancake are locking into a large spin  based on Ref.~\cite{aeppli2022hamiltonian, martin2013science}, which leads to the following 1D large-spin Hamiltonian:
\begin{equation}
    \begin{gathered}
    \hat{H}_{\mathrm{LS}}=\hat{H}_{\mathrm{on-site}}+\hat{H}_{\mathrm{off-site}},\\
    \hat{H}_{\mathrm{on-site}}/\hbar=\sum_n\Big[J_{0,n}\hat{\mathbf{S}}_n\cdot\hat{\mathbf{S}}_n+\chi_{0,n} \hat{S}_n^Z\hat{S}_n^Z+C_{0,n} \hat{N}_n \hat{S}^Z_n\Big],\\
    \hat{H}_{\mathrm{off-site}}/\hbar=\sum_n\Big[J_{1,n}\hat{\mathbf{S}}_n\cdot\hat{\mathbf{S}}_{n+1}+\chi_{1,n}\hat{S}_n^Z\hat{S}_{n+1}^Z+D_{1,n}(\hat{S}_n^X\hat{S}_{n+1}^Y-\hat{S}_n^Y\hat{S}_{n+1}^X)\Big].\\
    \end{gathered}
    \label{eq:largespin}
\end{equation}
The collective spin operators are defined as $\hat{\tilde{\mathbf{S}}}_{n}=\sum_{n_xn_y}\sum_{\alpha\beta=\{e,g\}}\hat{\tilde{c}}^{\dag}_{n_xn_yn,\alpha}\bm{\sigma}_{\alpha\beta}\hat{\tilde{c}}_{n_xn_yn,\beta}/2$ in the lab frame, where $\bm{\sigma}_{\alpha\beta}$ are Pauli matrices, $\hat{\tilde{c}}_{n_xn_yn,\alpha}$ are fermionic annihilation operators for radial mode labelled by $(n_x,n_y)$ assuming separable potential in pancakes, Wannier-Stark level $n$ along gravity and internal state $\alpha$. $\hat{N}_n$ is the atom number operator for Wannier-Stark level $n$.
We transform into the ``spiral'' frame by unitary transformation $\hat{ S}^{\pm}_{n}=e^{\pm i n\varphi}\hat{\tilde S}^{\pm}_{n}$ and $\hat{S}^{z}_{n}= \hat{\tilde S}^{z}_{n}$.
The interaction parameters are
\begin{equation}
    \begin{gathered}
    J_{0,n}=\eta_0(V_{eg}^{n,n}-U_{eg}^{n,n})/2, \quad \chi_{0,n}=\eta_0(V_{ee}^{n,n}+V_{gg}^{n,n}-2V_{eg}^{n,n})/2, 
    \\
    \quad C_{0,n}=\eta_0(V_{ee}^{n,n}-V_{gg}^{n,n})/2, \quad
    J_{1,n}=-\eta_1U_{eg}^{n,n+1}\cos\varphi, 
    \\
    \chi_{1,n}=-\eta_1U_{eg}^{n,n+1}(1-\cos\varphi), \quad D_{1,n}=-\eta_1U_{eg}^{n,n+1}\sin\varphi.\\
    \end{gathered}
\end{equation}
where $\varphi = 2 \pi a / \lambda_{clk}$ is the spin-orbit-coupled clock phase between nearest-neighbor sites of the lattice, with $a$ the lattice spacing.
$\eta_0$ and $\eta_1$ are dimensionless overlap integrals for on-site and nearest-neighbor interaction respectively,
\begin{equation}
    \eta_{|n-m|}=\frac{\sqrt{2\pi}}{k_L}\bigg(\frac{V_z}{E_{R}}\bigg)^{-1/4}\int\mathrm{d}z\,[W_n(z)]^2[W_m(z)]^2,
    \label{eq:integral}
\end{equation}
where $E_{R}=\hbar^2k_L^2/2m$ is the lattice recoil energy, with wave number $k_L=\pi/a$, and $W_n(z)$ is the wave function of a Wannier-Stark state centered at site $n$.

The $s$-wave ($U_{\alpha\beta}$) and $p$-wave ($V_{\alpha\beta}$) interaction strengths ($\alpha,\beta=\{g,e\}$) are calculated by averaging a Fermi-Dirac distribution over radial modes, 
\begin{equation}
    \begin{gathered}
    U_{\alpha\beta}^{n,m}=\frac{8\pi\hbar a_{\alpha\beta}}{m}\frac{k_L}{\sqrt{2\pi}}\bigg(\frac{V_z}{E_{R}}\bigg)^{1/4}\sum_{n_xm_xn_ym_y} s_{n_xm_x}s_{n_ym_y}\frac{N_{n_xn_yn}}{N_{n,\mathrm{init}}}\frac{N_{m_xm_ym}}{N_{m,\mathrm{init}}},\\ V_{\alpha\beta}^{n,m}=\frac{6\pi\hbar b_{\alpha\beta}^3}{m}\frac{k_L}{\sqrt{2\pi}}\bigg(\frac{V_z}{E_{R}}\bigg)^{1/4}\sum_{n_xm_xn_ym_y}(p_{n_xm_x}s_{n_ym_y}+s_{n_xm_x}p_{n_ym_y})\frac{N_{n_xn_yn}}{N_{n,\mathrm{init}}}\frac{N_{m_xm_ym}}{N_{m,\mathrm{init}}},
    \end{gathered}
    \label{eq:spd}
\end{equation}
where $a_{\alpha\beta}$ is the elastic $s$-wave scattering length, and $b_{\alpha\beta}^3$ is the elastic $p$-wave scattering volume. As the atoms are nuclear-spin polarized ($m_F=\pm9/2\rightarrow m_F'=\pm9/2$ transition between $^1S_0$ and $^3P_0$ states), to fully anti-symmeterize the wavefunction the following scattering lengths are required $a_{eg}=a^{-}_{eg}$, $b^3_{eg}=(b_{eg}^{+})^3$ \cite{aeppli2022hamiltonian,zhang2014spectroscopic}.
Defining the wave function for a radial mode ($n_x,n_y$) is $\phi_{n_x}\phi_{n_y}$, we have $s_{nm}=\int\mathrm{d}x\,[\phi_{n}(x)]^2[\phi_{m}(x)]^2$, $p_{nm}=\int\mathrm{d}x\,[(\partial_x\phi_{n}(x))\phi_{m}(x)-\phi_{n}(x)(\partial_x\phi_{m}(x))]^2$.
Here $N_{n_xn_yn}$ are the initial population in radial mode $(n_x,n_y)$ and lattice site $n$ under a Fermi-Dirac distribution, $N_{n_xn_yn}=\Big[\exp[(\epsilon_{n_xn_y}-\mu_n)/k_BT_R]+1\Big]^{-1}$, where the chemical potential for each lattice site $\mu_n$ is chosen to match the initial atom number for each Wannier-Stark level $N_{n,\mathrm{init}}=\sum_{n_xn_y}N_{n_xn_yn}$, and $T_R$ is the radial temperature. Errorbands in Fig.~\ref{fig:fig3} include the uncertainty of $s$-wave and $p$-wave scattering parameters,
$0.5 E_{R}$ uncertainty of lattice depth, as well as $20 \%$ uncertainty in radial temperature.

Apart from unitary evolution under $\hat{H}_{\mathrm{LS}}$, inelastic on-site p-wave $e$-$e$ collision can lead to two-body loss of the atom number as observed in previous studies \cite{martin2013science,zhang2014spectroscopic}.
We describe the atom loss based on the following Lindblad master equation,
\begin{equation}
    \hbar\frac{d}{dT}\hat{\rho}=-i[\hat{H}_{\mathrm{LS}},\hat{\rho}]+\sum_{n}\Gamma_{0,n}\mathcal{L}_n(\hat{\rho}),
    \label{eq:lindblad}
\end{equation}
where $\hat{H}_{\mathrm{LS}}$ is the Hamiltonian given in Eq.~(\ref{eq:largespin}).  The Liouvillian for $e$-$e$ loss is given by
\begin{equation}
    \mathcal{L}_n(\hat{\rho})=\sum_{n_xn_ym_xm_y}\bigg[\hat{L}_{n_xn_ym_xm_y}\hat{\rho}\hat{L}^{\dag}_{n_xn_ym_xm_y}-\frac{1}{2}\{\hat{L}^{\dag}_{n_xn_ym_xm_y}\hat{L}_{n_xn_ym_xm_y},\hat{\rho}\}\bigg],
\end{equation}
where $\hat{L}_{n_xn_ym_xm_y}=\hat{\tilde c}_{n_xn_yn,e}\hat{\tilde c}_{m_xm_yn,e}$. We use the averaged $e$-$e$ loss rate over the radial modes to maintain the large-spin description,
\begin{equation}
    \Gamma_{0,n}=\eta_0\tilde{V}^{n,n}_{ee}/4,
\end{equation}
where we replace the elastic $p$-wave scattering volume $b_{ee}^3$ in $V_{ee}^{n,m}$ by inelastic $p$-wave scattering volume $\beta_{ee}^3$ to get $\tilde{V}_{ee}^{n,m}$. For simplicity, we assume $U_{\alpha\beta}^{n,m}$, $V_{\alpha\beta}^{n,m}$ and $\tilde{V}_{ee}^{n,m}$ approximately unchanged under atom loss.
Due to the XY8 pulse sequence, one can assume the atom loss for ground and excited states is symmetric, and obtain the following equation for atom loss,
\begin{equation}
    \frac{d}{dT}N_n=-\frac{\Gamma_{0,n}}{2}N^2_n,
\end{equation}
which gives an exact solution
\begin{equation}
    N_n(T)=\frac{N_{n,\mathrm{init}}}{1+\Gamma_{0,n}N_{n,\mathrm{init}}T/2}.
\end{equation}
We fit the total atom number measured in the experiment integrating through all lattice layers with the fitting function $A/(1+BT)$ using fitting parameters $A,B$ to extract the atom loss rate, and then compare with the analytic solution above.

We perform numerical simulation based on truncated Wigner approximation (TWA) \cite{polkovnikov2010phase}. The key idea is to solve the mean-field equations of Eq.~(\ref{eq:lindblad}) with random sampling of initial conditions. 
For the initial state (``spiral'' frame) with all the spins pointing towards $+X$ direction, we set $S^X_n(0)=N_{n,\mathrm{init}}/2$, $N_n(0)=N_{n,\mathrm{init}}$, and sample $S^Y_n(0)$ and $S^Z_n(0)$ using a Gaussian distribution $\mathcal{N}(\mu=0,\sigma^2=N_{n,\mathrm{init}}/2)$.

In the case of $V_{\perp}=0$, we consider the radial modes as harmonic oscillator modes with trapping frequency $\omega_R=\sqrt{4V_z/mw_L^2}$, where $w_L$ is the Gaussian beam waist of the vertical lattice. We determine the radial temperature $T_R$ by comparing the density distribution projected to the x-y plane between theory and experiment at $17.4E_{R}$, which leads to $T_R=250$nK at this lattice depth. Since the lattice depth is ramping up adiabatically, the ratio $k_BT_R/\hbar\omega_R$ should be roughly a constant, we use $T_R(\mathrm{nK})=60\times \sqrt{V_z/E_R}$ to generate atom distribution in radial modes.

In the case of $V_{\perp}>0$, the radial modes are generated by the potential of a 2D lattice with lattice depth $V_{\perp}$ plus harmonic oscillator with trapping frequency $\omega_R=\sqrt{4(V_z+V_{\perp})/mw_L^2}$, where we assume the Gaussian beam waist is nearly the same for all lattice beams, i.e.~$w_L\approx w$. The compression step in the loading sequence leads to a lower temperature compared to the case of $V_{\perp}=0$, and we use radial temperature $T_R (V_{\perp}>0) = 0.42 \times T_R (V_{\perp}=0)$ to generate atom distribution in radial modes.

The validity of the 1D spin model is based on the frozen-mode approximation \cite{aeppli2022hamiltonian, martin2013quantum}, which is to assume all the atoms are fixed in their single-particle eigenstates due to negligible effects of interaction on the single-particle energy spectrum, ensured by $N_{n,\mathrm{init}}J_{0,n}\ll \hbar\omega_R$ and $N_{n,\mathrm{init}}J_{0,n}\ll mga$.
We restrict our calculation within $V_{\perp}\leq 6E_R$ to avoid the breakdown of this approximation.

\subsection*{Spin wave analysis for 1D spin model}
Here we perform a spin wave analysis to $\hat{H}_{\mathrm{LS}}$ (see Eq.~(\ref{eq:largespin})) to have a further understanding of Ramsey contrast decay beyond numerical simulations. We assume the same atom number $N_s$ for each lattice site, periodic boundary conditions, and no atom loss. We drop the terms proportional to $\hat{S}^Z_n$ since it is suppressed by the XY8 pulses.
Considering the initial state with all the spins pointing to $+X$ direction, we perform a Holstein–Primakoff transformation to the large-spin operators, $\hat{S}_n^X = N_s/2-\hat{a}_n^{\dag}\hat{a}_n$, $\hat{S}_n^Y \approx \sqrt{N_s}(\hat{a}_n+\hat{a}_n^{\dag})/2$, $\hat{S}_n^Z \approx \sqrt{N_s}(\hat{a}_n-\hat{a}_n^{\dag})/(2i)$.
In this way, the initial state becomes the vacuum state of the bosonic operators. We keep the terms up to quadratic order of bosonic operators, and then apply a Fourier transform to obtain the bosonic operators for spin waves ($k\in(-\pi,\pi]$), $\hat{a}_n=\sum_k e^{ikn}\hat{a}_k/\sqrt{L}$, with $L$ the number of lattice sites. We get
\begin{equation}
    \hat{H}_{\mathrm{LS}} \approx \frac{1}{2}\sum_k c_1 (\hat{a}_k^{\dag}\hat{a}_k+\hat{a}_k^{\dag}\hat{a}_k)-c_2(\hat{a}_k^{\dag}\hat{a}_{-k}^{\dag}+\hat{a}_{-k}\hat{a}_k),
\end{equation}
where
\begin{equation}
    c_1 =c_2-N_sJ_1\Big(1-\cos(k)\Big), \quad c_2= \frac{N_s\chi_0}{2}+\frac{N_s\chi_1}{2}\cos(k).
\end{equation}
Here, $J_1$, $\chi_0$, $\chi_1$ are the mean values of $J_{1,n}$, $\chi_{0,n}$, $\chi_{1,n}$. So the excitation numbers for spin waves is given by
\begin{equation}
    n_{\pm k}(T)=\langle\mathrm{vac}|\hat{a}^{\dag}_k(T)\hat{a}_k(T)|\mathrm{vac}\rangle=\langle\mathrm{vac}|\hat{a}^{\dag}_{-k}(T)\hat{a}_{-k}(T)|\mathrm{vac}\rangle=c_2^2\Bigg[\frac{\sin\Big(T\sqrt{c_1^2-|c_2|^2}\Big)}{\sqrt{c_1^2-|c_2|^2}}\Bigg]^2.
\end{equation}
We can express the Ramsey contrast at short time in terms of the excitation numbers for spin waves,
\begin{equation}
    C(T)=1-\frac{2}{N_sL}\sum_kn_k(T).
\end{equation}

Now we discuss the physics of spin-wave excitations for spin-wave modes $k=0$ (minimizing the difference between $c_1$ and $c_2$) and $k=\pi$ (maximizing the difference between $c_1$ and $c_2$). In the case of $k=0$, we have $n_{k=0}=N_s^2(\chi_0+\chi_1)^2T^2/4$, which leads to quadratic growth of the $k=0$ mode. In the case of $k=\pi$, we have $n_{k=\pi}=N_s^2(\chi_0-\chi_1)^2\sinh^2(KT)/(4K^2)$, where $K=N_s\sqrt{2J_1(\chi_0-\chi_1-2J_1)}$. Experimental values of the interaction strengths ensure $K$ is a real number. It seems we always have exponential growth of the $k=\pi$ mode. However, the spin wave analysis breaks down for $n_k\sim N_s$. If $K$ is small enough such that we reach $n_{k=\pi}\sim N_s$ at a time $KT\ll 1$, we can approximate $n_{k=\pi}\approx N_s^2(\chi_0-\chi_1)^2T^2/4$, which returns to quadratic growth. Therefore, exponential growth only occurs in the regime $N_s(\chi_0-\chi_1)^2 \ll K^2$.

Based on the discussions above, we can separate the system dynamics considering quadratic or exponential growth for the $k=\pi$ mode. 
As discussed in \cite{aeppli2022hamiltonian}, $s$-wave and $p$-wave interaction strengths in the 1D lattice have different dependence on the lattice depth $V_z$. When increasing $V_z$, $p$-wave interaction strength increases,  $\chi_0\propto V_z^{5/4}$, while $s$-wave interaction strength decreases, $J_1, \chi_1\propto e^{-4\sqrt{V_z/E_R}}$.
So we can access these two regimes at different $V_z$.

In the regime of large $V_z$, the system dynamics can be described by quadratic growth for all the spin wave modes. So we obtain the Ramsey contrast by summing over all the spin wave modes, $C(T)=1-N_s(\chi_0^2+\chi_1^2/2)T^2/2$, which agrees with the short-time expansion under Ising-type interactions by setting $J_1=0$. We can estimate the $T_2$ coherence time for quadratic growth,
\begin{equation} \label{eq:spinwave_t2q}
    T_2^q \sim \frac{1}{\sqrt{N_s}}\frac{1}{\sqrt{\chi_0^2+\chi_1^2/2}}.
\end{equation}
Since this regime is mainly dominated by $p$-wave interactions, the coherence time decreases as we increase $V_z$ (see Fig.~S5).

In the regime of small $V_z$, the system dynamics is dominated by exponential growth near the $k=\pi$ mode. To obtain the Ramsey contrast we only consider the spin wave modes near $k=\pi$, which leads to $1-C(T)\sim N_s (\chi_0-\chi_1)^2e^{2KT}/K^2$. We can estimate the $T_2$ coherence time for exponential growth,
\begin{equation} \label{eq:spinwave_t2e}
    T_2^e \sim \frac{1}{K} = \frac{1}{N_s} \frac{1}{\sqrt{2J_1(\chi_0-\chi_1-2J_1)}}.
\end{equation}
Since this regime requires the existence of $s$-wave interaction, the coherence time increases as we increase $V_z$ (see Fig.~S5).

\begin{figure*}[hbtp]
    \centering
    \includegraphics[width=5in]{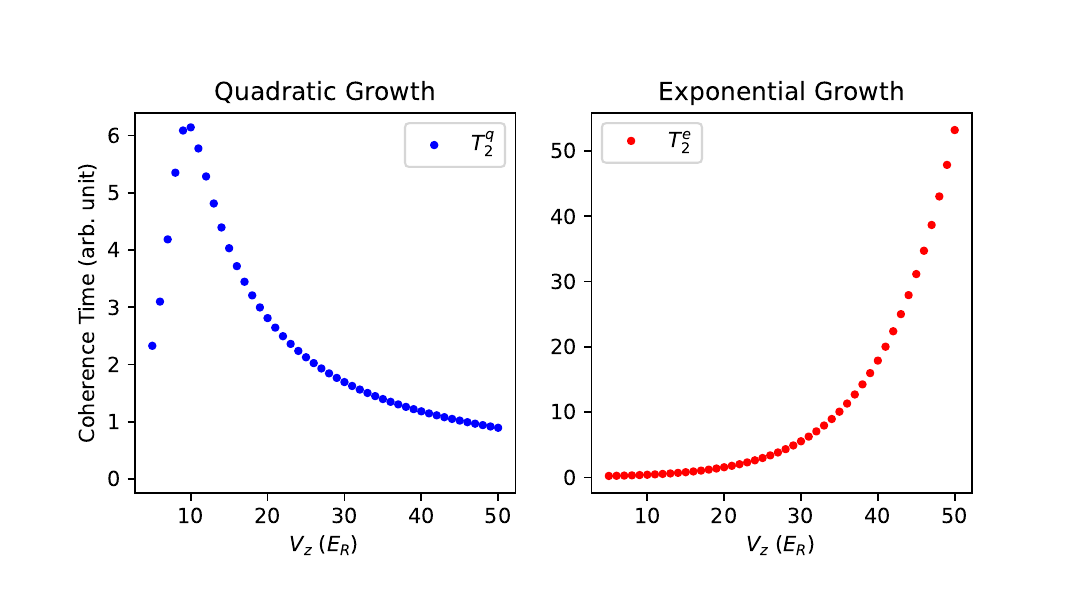}
    \caption{
	    \textbf{Impact of spin wave growth on coherence times in the 1D limit}. The plots show the qualitative impact of the quadratically and exponentially growing spin wave modes on the coherence time $T_2$ according to Eqs.~(\ref{eq:spinwave_t2q}) and (\ref{eq:spinwave_t2e}). At large lattice depths the $T_2$ time is limited by the quadratically growing modes while the coherence in a shallow lattice is dominated by the exponential instability around $k=\pi$. Here we present the coherence time in arbitrary unit since Eqs.~(\ref{eq:spinwave_t2q}) and (\ref{eq:spinwave_t2e}) are only order-of-magnitude estimations up to unknown constant factors of order 1.}
    \label{fig:figs5}
\end{figure*}

The crossover between these two regimes occurs when $T_2^q$ is comparable with $T_2^e$. Due to the different dependence of the atom number per site $N_s$ in these two regimes, the crossover point depends on $N_s$, i.e. $N_s J_1/\chi_0 \sim 1$. Note that at the crossover point the $p$-wave interaction strength is already much larger than the $s$-wave interaction strength. This explains why the crossover point occurs at a larger lattice depth $V_z$ compared to the cancellation of density shift in \cite{aeppli2022hamiltonian}.


\subsection*{Full simulation of the 3D cloud}

To model the experimental contrast decay in Fig.~\subref{fig:fig4}{A} and Fig.~\subref{fig:fig4}{B}, we average over the 3D distribution using calibrated experimental parameters. We first consider the initialization of the atomic cloud in the 3D lattice + XODT trap. Assuming that the XODT trap depth is equal along the $x$, $y$ is equal, the system has cylindrical symmetry in the $x - y$ plane, and all vertical $z$-direction tubes at fixed horizontal radius $r_{xy} = \sqrt{x^2 + y^2}$ from the trap center have the same physics.

Within a given lattice tube at fixed $r_{xy}$, we assume that the loading procedure yields an ensemble of atoms in a discrete Fermi-Dirac distribution,
\begin{equation}
\begin{aligned}
P(r_{xy},z) &= \frac{1}{e^{(E(r_{xy},z)-\mu)/(k_B T_{lat})}+1},\\
E(r_{xy},z) &= \frac{1}{2}m \omega_{XODT,\perp}^2 a^2 r_{xy}^2 + \frac{1}{2}m \omega_{XODT}^2 a^2 z^2,
\end{aligned}
\end{equation}
where $z$ is the integer lattice site index along the vertical tube direction, $T_{lat}$ is the temperature in the lowest band of the 3D lattice, $k_B$ is Boltzmann's constant, $\mu$ is a chemical potential offset, and $E(r_{xy},z)$ is the local potential energy from the XODT, with $(\omega_{XODT,\perp}, \omega_{XODT})/(2\pi)=(136,250)$ Hz the horizontal and vertical trapping frequencies of the XODT (the actual XODT trap in the experiment has slightly unequal trapping frequencies along the $x$ and $y$ horizontal directions, so we take $
\omega_{XODT,\perp}$ as an average of the two for simplicity). This approach employs the local density approximation, neglecting any tunneling effects or band excitations. We omit gravity by assuming that it is incorporated into the definition of the $z$-direction potential; i.e. the minimum of the XODT potential and gravity is at $z= 0$, and the overall center of the potential is at $r_{xy}=z=0$ during the loading process.

The total number of atoms is given by the integral of the probability density,
\begin{equation}
N =  \int_{0}^{\infty}d r_{
xy}\> r_{xy}\int_{-\infty}^{\infty}\> d z\> P(r_{xy},z) = - \left(\frac{2\pi k_B T_{lat}}{m a^2}\right)^{3/2} \frac{1}{\omega_{XODT,\perp}^2 \omega_{XODT}} \text{Li}_{3/2} \left[-e^{\mu/(k_B T_{lat})}\right],
\end{equation}
where $\text{Li}_{n}[.]$ is the polylogarithm function. For fixed trapping frequencies and atom number $N$, the gas is described by just the temperature $T_{lat}$, while the chemical potential offset $\mu$ is fixed via the above equation. To account for different possible initial temperatures of the gas, in our analysis we fix the atom number at $N = 30,000$, and vary $T_{lat}$.

After initialization, the Ramsey dynamics are modeled as follows. For each horizontal trap radius, $r_{xy}$, a set of initial particle distributions is randomly sampled via the Fermi-Dirac distribution. Each such distribution is a set of occupied sites $\mathbbm{P} = \{z\}$. From this sampling we identify all continuous, uninterrupted chains of atoms  For example, if we sampled particles at sites $\mathbbm{P} = \{-3,-2,-1,4,5,9\}$, we identify uninterrupted chains $(-3,-2,-1),\> (4,5),\>(9)$, assuming that atoms separated by one or more empty sites do not talk to each other.

For each uninterrupted chain, the quantum dynamics are modelled following the description in the $\textit{Superexchange contrast dynamics}$ section of the Supplementary Materials. The sampling from the Fermi-Dirac distribution is centered about the XODT potential minimum $j_s$, while the local energy differences $\Delta E_{j}$ that enter into the superexchange couplings are determined by the lattice potential minimum $j_0$ via $\Delta E_{j} = \frac{1}{2} \omega_{lat}^2 a^2 [(j+1-j_0)^2 - (j-j_0)^2]$. Note that to avoid numerical issues in places where the superexchange denominator $U - \Delta E_{j}$ vanishes, we also bound the magnitude of $J_{\textrm{SE}}(j)$ by the tunneling rate $t_{z}$. While there can be beyond-superexchange physics in such places, we assume such effects are localized to a few sites and do not affect the overall signal.

The contributions of each uninterrupted chain in the set $\mathbbm{P}$ to the final excited state fraction as a function of final Ramsey pulse angle $\theta$ are solved separately, then summed together,
\begin{equation}
N_{r_{\mathrm{XY}},\theta}(T) = \sum_{\mathbbm{P}} N_{\theta}(T).
\end{equation}
Many thermal trajectories with different initial particle distributions $\mathbbm{P}$ are sampled, and an average is taken for each time step. This calculation is repeated for discrete horizontal trap radii $r_{\mathrm{XY}}= 0 \dots 16$ sites corresponding to a horizontal cylinder of roughly $6 \mu$m, which is the experimentally imaged region $P_1$. The contributions of tubes at different $r_{xy}$ are then averaged, weighted by the circumference $2\pi r_{xy}$, which is the number of tubes at that radius,
\begin{equation}
N_{\mathrm{total},\theta}(T) = \sum_{r_{xy}} (2 \pi r_{xy}) N_{r_{xy},\theta}(T).
\end{equation}
The total contrast is then obtained via,
\begin{equation}
C(T) = \mathcal{N}\left[\text{max}_{\theta} N_{\mathrm{total},\theta}(T) - \text{max}_{\theta} N_{\mathrm{total},\theta}(T)\right],
\end{equation}
where $\mathcal{N}$ is a normalization prefactor chosen to fix $C=1$ at $T=0$.

This calculation is repeated for a range of different initial temperatures $T_{lat}$. Higher temperatures lead to a larger long-time average value for the contrast, but do not strongly affect the oscillatory profile. This is because the temperature only determines how likely one is to find atoms next to each other in longer chains, but does not directly change the ensuing quantum dynamics of the chains. The main text Figs.~\subref{fig:fig4}{A} and ~\subref{fig:fig4}{B} numerical results use temperatures of $570$ nK and $380$ nK respectively. Note that this is not the true temperature, since we fix atom number $N=30,000$ in the simulations for simplicity, which was not directly controlled in the experiment. However this temperature is within the same order of magnitude as that predicted for the same experimental platform in earlier work.

\clearpage
\bibliography{apssamp}
\bibliographystyle{Science}

\end{document}